\documentclass[5p,twocolumn]{elsarticle}

\usepackage{etex,pstricks,graphicx,amsmath,bbm,mathrsfs,amssymb,psfrag,mathabx,times}
\usepackage[colorlinks=true,citecolor=blue,linkcolor=blue,urlcolor=blue]{hyperref}
\usepackage[english]{babel}

\def\bmE{{\mathbf E}}
\def\bmr{{\mathbf r}}
\def\bmp{{\mathbf p}}
\def\e{\text{e}}
\def\hq{\hat{q}}
\def\hp{\hat{p}}
\def\hH{\hat{H}}
\def\hI{\hat{I}}
\def\ha{\hat{a}}
\def\hb{\hat{b}}
\def\hc{\hat{c}}
\def\hd{\hat{d}}
\def\hx{\hat{x}}
\def\hy{\hat{y}}
\def\hn{\hat{n}}
\def\hA{\hat{A}}
\def\hB{\hat{B}}
\def\hD{\hat{D}}
\def\hU{\hat{U}}
\def\hrho{\hat{\varrho}}
\def\d{\mbox{d}}
\def\hJ{\hat{J}}
\def\hS{\hat{S}}
\def\hT{\hat{T}}
\def\hK{\hat{K}}
\def\cE{{\mathcal E}}
\def\cS{{\mathcal S}}
\def\cI{{\mathcal I}}
\def\cF{{\mathcal F}}
\def\cQ{{\mathcal Q}}

\newcommand{\rmpl}[1]{\mbox{\footnotesize #1}}

\def\Ubs{\hat{U}_{\rmpl{BS}}}

\def\idop{\hat{\mathbb I}}
\def\Real{\Re\mbox{e}}

\newcommand{\ket}[1]{\vert #1 \rangle}
\newcommand{\bra}[1]{\langle #1 \vert}

\newcommand{\braket}[2]{\langle #1 \vert #2 \rangle}

\makeatletter 
\newcommand*\superimpose[2]{%
  \ooalign{$\m@th#1\@firstoftwo#2$\cr
           \hidewidth$\m@th#1\@secondoftwo#2$\hidewidth}%
}
\makeatother 

\newcommand*\threedotsord{\mathpalette\superimpose{{\mathop:}{\cdot}}} 
\newcommand*\threedotsbin{\mathbin{\threedotsord}}     

\newtheorem{thm}{Theorem}
\newtheorem{exercise}{$\square$ Exercise}
\newtheorem{example}{$\square$ Example}

\begin{document}

\title{Introduction to generation, manipulation and characterization of optical quantum states}

\author{Stefano Olivares\corref{cor1}}%
\ead{stefano.olivares@fisica.unimi.it}
\address{Dipartimento di Fisica ``Aldo Pontremoli'', Universit\`a degli Studi di Milano, via~Celoria~16, I-20133 Milan, Italy}
\address{INFN, Sezione di Milano, I-20133 Milan, Italy}

\cortext[cor1]{Corresponding author}


\begin{abstract}
In this brief tutorial we provide the theoretical tools needed to describe the generation, manipulation and characterization of optical quantum states and of the main passive (beam splitters) and active (squeezers) devices involved in experiments, such as the Hong-Ou-Mandel interferometer and the continuous-variable quantum teleportation. We also introduce the concept of operator ordering and the description of a system by means of the $p$-ordered characteristic functions. Then we focus on the quasi-probability distributions and, in particular, on the relation between the marginals of the Wigner function and the outcomes of the quadrature operator measurement. Finally, we introduce the balanced homodyne detection to measure the quadrature operator and the homodyne tomography as a tool for characterizing quantum optical states also in the presence of non-unit quantum efficiency.
\end{abstract}

\maketitle

\section{Introduction}

The development of optics is focused to understand and interpret the many optical phenomena observed (interference, diffraction,$\ldots$). In the 19$^{\rmpl{th}}$ century the investigation of these phenomena was restricted to direct observation and photographic plates. The information available was about the spatial distribution of light, being restricted to averages, and there was a lack of knowledge about the time resolved information. In order to describe the experimental results, only the geometrical optics (coming from Greeks) and the wave optics (due to Huygens' ideas and Fresnel's mathematical models) were enough. When the laser was invented (and experimentally realized) it became possible to fully exploit the coherence properties of light (holography experiments), to perform the spectroscopic investigation of the (quantum) atom, the dynamics of chemical and biological processes and so on. Furthermore, the high intensities achievable with laser beams made possible the realization of nonlinear optical effects (generation of sum-, difference- and high-order frequencies) \cite{bachor}.
\par
Optics is one of the best testing grounds of quantum mechanics. Today, the investigation of optics, or, rather, quantum optics is based on {\it lasers}, as light sources, and on the {\it photoelectric effect}, as detection strategy: in this way it is also possible to generate exotic quantum states.
It is worth noting that, by means of the photoelectric detectors, one can address individual events which generate photocurrents. All this unavoidably leads to the advent of the so-called statistical optics.
\par
The following pages are not a review paper on ``quantum optics'' and, therefore, we have chosen to cite only few necessary references that the reader can use to improve the study. This is just a tutorial, with useful examples and simple exercises. It aims at introducing the reader to the (not always) basic elements of quantum optics, useful to quickly start to deal with it. The advanced student will find a useful compendium of the main quantum optics elements, while the beginner will meet the most common optical quantum states (Fock, coherent, thermal states), quantum optical devices (beam splitters, squeezers) and the fundamental tools to describe them (operators ordering, quasi-probability distributions, homodyne tomography).
\par
The plan of the tutorial is the following. After reviewing the mathematical description of the classical monochromatic  wave (sect.~\ref{s:cl:wave}), we discuss the quantization of the electromagnetic field and its basic states (sect.~\ref{s:quantization}), also providing the fundamental tools to deal with their time evolution (sect.~\ref{s:theorems}). The reader will meet the theoretical description of the linear and bilinear interactions of modes usually exploited in quantum optics: the beam splitter implementing the mode-mixing interaction (sect.~\ref{s:BS}), the single-mode (sect.~\ref{s:smSq}) and the two-mode squeezing operations (sect.~\ref{s:tmSq}).
Then, we turn the attention on the characterization of the states.
A method to study the photon number statistics based on the moment generating functions is provided (sect.~\ref{s:pn}). In order to describe the properties of the quantum states, we introduce the Fano factor and the Mandel parameter (sect.~\ref{s:fano}) and the operator ordering (sect.~\ref{s:ordering}).
\par
The last part of the tutorial is devoted to a different approach to describe quantum states, based on the characteristic functions (sect.~\ref{s:char}) and the quasi-probability distributions (sect.~\ref{s:QP}).
This approach leads us to the concept of nonclassicality related to the phase-space analysis and we also briefly discuss the main techniques to generate nonclassical optical states (sect.~\ref{s:NCstates}).
Thereafter, we show the link between the Wigner function and the expectation values of the quadrature operators (sect.~\ref{s:Wexp}) and how it is possible to measure the expectations of quadrature operators through the so-called homodyne detection (sect.~\ref{s:QOexp}). The working principle of the homodyne tomography is also explained in some detail (sect.~\ref{s:tomo}). Eventually, we close the tutorial drawing some concluding remark (sect.~\ref{s:concl}).

\section{Classical waves and quadratures}\label{s:cl:wave}
An electromagnetic wave in isotropic insulating medium is described by the wave equation:
\begin{equation}
\nabla^2\bmE(\bmr,t) -\frac{1}{c^2}\frac{\partial^2}{\partial t^2} \bmE(\bmr,t)=0
\end{equation}
whose (monochromatic) solution can be written as:
\begin{equation}\label{class:wave}
\bmE(\bmr,t) =\left[
\alpha(\bmr)\,\e^{-i\omega t} + \alpha^*(\bmr)\,\e^{i\omega t}
\right]\, \bmp(\bmr),
\end{equation}
where $\bmr$ and $t$ represent the position vector and the time, respectively, $\alpha(\bmr)$ is a complex amplitude function, $\omega$ is the frequency of the wave and $\bmp(\bmr)$ the polarization vector. We can rewrite the amplitude $\alpha(\bmr)$ as:
\begin{equation}
\alpha(\bmr) = \alpha_0(\bmr)\, \e^{i\phi(\bmr)},
\end{equation}
where $\alpha_0(\bmr)$ is the (dimensional) magnitude of the field and the phase term $\phi(\bmr)$ determines the shape of the wave front. It is worth noting that the spatial distribution of $\phi(\bmr)$ describes the curvature of the wave. In the case of a \emph{plane wave} with wave vector $k=\omega/c$ and moving along the positive direction of $z$-axis we have:
\begin{equation}
\alpha(\bmr) = \alpha_0 \, \e^{i k z} \Rightarrow
\bmE(\bmr,t) = 2 \alpha_0 \cos (k z-\omega t)\, \, \bmp(\bmr).
\end{equation}
\par
\begin{figure}[t]
\begin{center}
    \includegraphics[width=0.8\columnwidth]{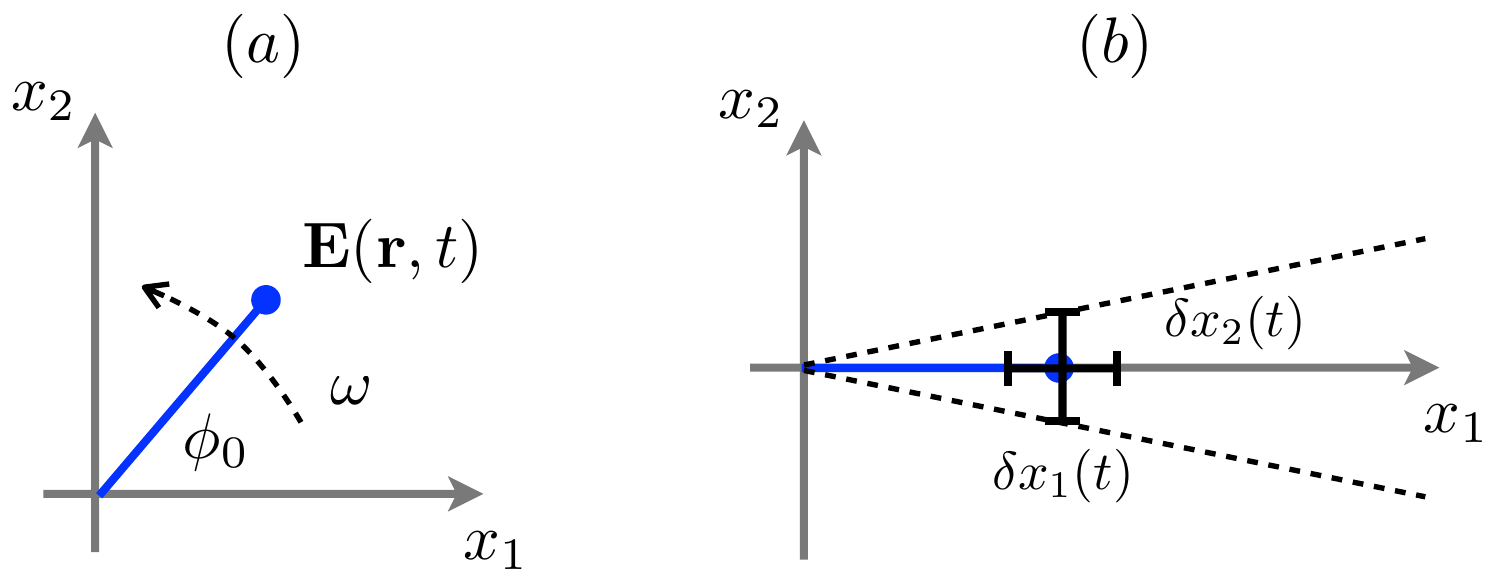}
\caption{\label{f:phasor} (a) Phasor representation of the classical wave in Eq.~(\ref{class:wave:1}). (b) In the limit $\alpha_0 \gg \delta x_1(t)$ the fluctuations of the quadratures $\delta x_1(t)$ and $\delta x_2(t)$ correspond to amplitude and phase fluctuations (see the text for details).}
\end{center}
\end{figure}

In general, the phase function $\phi(\bmr)$ describes the shape of the wave front as well as its absolute phase with respect to a reference. If we introduce the following \emph{quadratures}:
\begin{subequations}\label{class:quad}
\begin{align}
x_1(\bmr) &= \alpha_0(\bmr) + \alpha_0^*(\bmr), \\[1ex]
x_2(\bmr) &= -i\left[\alpha_0(\bmr) - \alpha_0^*(\bmr)\right],
\end{align}
\end{subequations}
we can explicitly write the absolute phase as:
\begin{equation}
\phi_0(\bmr) = \tan^{-1}\left[ \frac{x_2(\bmr)}{x_1(\bmr)} \right],
\end{equation}
and Eq.~(\ref{class:wave}) rewrites:
\begin{equation}\label{class:wave:1}
\bmE(\bmr,t) = E_0 \left[
x_1(\bmr)\,\cos \omega t + x_2(\bmr)\,\sin \omega t
\right]\, \bmp(\bmr).
\end{equation}
In Fig.~\ref{f:phasor}~(a) we show the graphical representation of the classical wave in Eq.~(\ref{class:wave:1}) in a given space point $\bmr$ at time $t$, which corresponds to a point in the two-dimensional $x_1$--$x_2$ plane. This kind of representation of the classical fields can be used to describe interference and diffraction phenomena.
\par
In the presence of quadrature fluctuations, the complex amplitude $\alpha(t)$ of a classical field can be written as (without loss of generality we assume $\alpha_0 \in {\mathbb R}$):
\begin{equation}
\alpha(t) = \alpha_0 + \delta x_1(t) + i \delta x_2(t).
\end{equation}
In the limit $\alpha_0 \gg \delta x_1(t), \delta x_2(t)$, it is clear from Fig.~\ref{f:phasor}~(b) that $\delta x_1(t)$ leads to fluctuations of the amplitude of $\alpha_0$, whereas $\delta x_2(t)$ is responsible of its phase fluctuations. Therefore, one usually refers to the quadrature $x_1$ as ``amplitude'' and to $x_2$ as ``phase''. We recall that phase and amplitude modulation can be used to encode signals into a classical electromagnetic wave.
\par
Starting form the quadratures introduced in Eqs.~(\ref{class:quad}), we can define the generic quadrature (since we are considering a particular point of the space at a given time, we drop the explicit dependence on $\bmr$ and $t$):
\begin{equation}\label{x:theta:class}
x_\theta = x_1\, \cos\theta + x_2\, \sin\theta.
\end{equation}
In the presence of a classical field, the value of $x_\theta$ oscillates as a function of the quadrature phase $\theta$.

\section{Basics of electromagnetic field quantization}\label{s:quantization}
Modern light detectors are based on the photoelectric effect and, therefore, the introduction of the quantization of electromagnetic field is somehow natural. However, quantization is needed also to describe our vision process. Let us think of a starry night: if the vision process were based on a classical effect, the image of the stars would require many seconds to build up, that is the time needed to collect enough energy, like a 19$^{\rmpl{th}}$ century photographic plate. Since this is not the case (we see the faint stars!) and today we know that we need less than ten photons (depending on the frequency) to ``detect'' an object, we can conclude that a quantized description of light is needed.
\par
In this section we briefly review the main steps to write the quantized Hamiltonian of the electromagnetic field \cite{GK}. We consider a cavity with a roundtrip $L$, thus only the discrete optical modes whose frequencies satisfy the relation $\omega_l = 2 \pi c \, l / L$ , with $l = 0,1,2,\ldots$, are allowed. Upon introducing the quadratures $x_{1,l}$ and $x_{2,l}$ of each mode $l$, the Hamiltonian $H$ of the classical fields can be written as (for the sake of simplicity we introduce the quantum constant $\hbar$):
\begin{equation}
H = \sum_l \frac{\hbar \omega_l}{4}\left(
x_{1,l}^2+x_{2,l}^2 \right).
\end{equation}
In order to quantize $H$, we define the canonical variables:
\begin{equation}
q_l = \sqrt{\frac{\hbar}{2\omega_l}}\,x_{1,l} \quad \text{and} \quad
p_l = \sqrt{\frac{\hbar\omega_l}{2}}\,x_{2,l}
\end{equation}
and the previous Hamiltonian becomes:
\begin{equation}
H = \frac12 \sum_l \left( p_l^2 + \omega_l^2 q_l^2 \right).
\end{equation}
The canonical quantization is obtained substituting to $q_l$ and $p_l$ the canonical operators $\hq_l$ and $\hp_l$, respectively, with commutation relation $[\hq_l,\hp_k] = i \hbar \, \delta_{l,k}$, that is:
\begin{equation}
\hH = \frac12 \sum_l \left( \hp_l^2 + \omega_l^2 \hq_l^2 \right).
\end{equation}
We now introduce the annihilation and creation operators for the $l$-th mode, $\ha_l$ and $\ha^\dag_l$, respectively, with $[\ha_l,\ha_k^\dag] = \delta_{l,k}$, and define the following \emph{quadrature operators}:
\begin{equation}
\hx_{1,l} = \ha_l + \ha^\dag_l \quad \mbox{and} \quad
\hx_{2,l} = -i \left( \ha_l - \ha^\dag_l \right),
\end{equation}
and the position- and momentum-like operators:
\begin{subequations}
\begin{align}
\hq_l &= \sqrt{\frac{\hbar}{2\omega_l}}\,\left( \ha_l + \ha^\dag_l \right) \\[1ex]
\hp_l &= -i\sqrt{\frac{\hbar\omega_l}{2}}\,\left( \ha_l - \ha^\dag_l \right),
\end{align}
\end{subequations}
where $[\hx_{1,l},\hx_{2,k}] = 2 i \, \delta_{l,k}$. The quantum Hamiltonian $\hH$ can be written as:
\begin{equation}
\hH = \sum_l \hbar \omega_l \left( \ha_l^\dag \ha_l +\frac12 \right),
\end{equation}
that is the Hamiltonian of a set of {\it quantum harmonic oscillators} with frequencies $\omega_l$. From now on we consider a single mode with frequency $\omega$ described by the field operator $\ha$.
\par
As in the classical case, we can define the generic quadrature operator as:
\begin{align}\label{x:theta:quant}
\hx_\theta &= \hx_1\, \cos\theta + \hx_2\, \sin\theta \\[1ex]
&=\ha\, \e^{-i \theta} + \ha^{\dag}\, \e^{i \theta}.
\end{align}
We will see in the following that the expectation values of $\hx_\theta$ can be rather different with respect to the case of the classical quadrature.

\subsection{Fock states}
The eigenstates of the number operator $\hn = \ha^{\dag}\ha$ are the Fock states or number states $\ket{n}$, $n \in {\mathbbm N}$, and the corresponding eigenvalues are the integer numbers $n$, namely:
\begin{equation}
\hn \ket{n} = n \ket{n}.
\end{equation}
We recall that the Fock states are a resolution of the identity operator, $\sum_n \ket{n}\bra{n} = \idop$. The action of the annihilation and creation operators on a Fock state is:
\begin{subequations}
\begin{align}
\ha \ket{n} &= \sqrt{n}\ket{n-1},\\[1ex]
\ha^{\dag} \ket{n} &= \sqrt{n+1}\ket{n+1},
\end{align}
\end{subequations}
respectively. It is worth noting that the \emph{vacuum state} $\ket{0}$ is an eigenvector of the annihilation operator with zero eigenvalue. Sometimes it is useful to write $\ket{n}$ as a power of the creation operator applied to the vacuum state, namely:
\begin{equation}\label{n:a:dag}
\ket{n} = \frac{\left(\ha^{\dag}\right)^n}{\sqrt{n}} \ket{0}.
\end{equation}

\begin{exercise}
Calculate the variance $\Delta^2(\hn)$ of the number operator $\hn$ given the Fock state $\ket{m}$, namely:
$$\Delta^2(\hn) = \bra{m}\hn^2\ket{m} - \bra{m}\hn\ket{m}^2.$$
$ $ \hfill $\blacksquare$
\end{exercise}

Let's turn our attention on the expectations of the quadrature operator $\hx_\theta$. In the presence of a Fock state $\ket{n}$ it is easy to show that $\bra{n} \hx_\theta \ket{n} = 0$, $\forall n$ and $\forall \theta$. If we compare this result with the classical case, we can conclude that the number states are rather exotic states of light!

\subsection{Coherent states}
Light states with more familiar behavior are the eigenvectors of the annihilation operator:
\begin{equation}
\ha \ket{\alpha} = \alpha \ket{\alpha},\quad \alpha\in {\mathbbm C}.
\end{equation}
The states $\ket{\alpha}$ are called \emph{coherent states} and are the closest approximation of the output state of a laser.
\par
Exploiting the completeness relation $\sum_n\ket{n}\bra{n} = \idop$ and the normalization condition $\braket{\alpha}{\alpha}=1$ we can find the photon number expansion of a coherent state (see Example~\ref{ex:ch:st:pn}).

\begin{example}\label{ex:ch:st:pn}
In this example we calculate the photon number statistics $p(n)= |\braket{n}{\alpha}|^2$ of a coherent state $\ket{\alpha}$.
We start using the completeness relation of Fock states:
$$\ket{\alpha} \to \left( \sum_{n=0}^{\infty}\ket{n}\bra{n} \right) \ket{\alpha} = 
\sum_{n=0}^{\infty} \braket{n}{\alpha} \ket{n}.$$
Now we have:
$$\braket{n}{\alpha} = \bra{0}\frac{\left( \ha \right)^n}{\sqrt{n!}} \ket{\alpha} = 
\braket{0}{\alpha}\, \frac{\alpha^n}{\sqrt{n!}},$$
where $\braket{0}{\alpha} \in {\mathbbm C}$ and we used Eq.~(\ref{n:a:dag}).
Therefore, we can write:
$$\ket{\alpha} = \braket{0}{\alpha} \sum_{n=0}^{\infty}  \frac{\alpha^n}{\sqrt{n!}} \ket{n}.$$
In order to find an explicit expression of $\braket{0}{\alpha}$ we recall that the normalization condition requires $\braket{\alpha}{\alpha}=1$, namely:
$$1 = \braket{\alpha}{\alpha} = |\braket{0}{\alpha}|^2 \sum_{n=0}^{\infty} \frac{|\alpha|^{2n}}{n!}
= |\braket{0}{\alpha}|^2\, \e^{|\alpha|^2}
$$
and we obtain:
$$
|\braket{0}{\alpha}| =
\e^{-|\alpha|^2/2}.$$
In general $\braket{0}{\alpha}$ is a complex number, however a quantum state is defined up to an overall phase, thus we can put $\braket{0}{\alpha} = \e^{-|\alpha|^2/2}$. Finally we have:
\begin{equation}
\ket{\alpha} = \e^{-|\alpha|^2/2} \sum_{n=0}^{\infty} \frac{\alpha^n}{\sqrt{n!}}\ket{n},
\end{equation}
and the photon number distribution reads:
\begin{equation}
p(n) = |\braket{n}{\alpha}|^2 = \e^{-|\alpha|^2}\,\frac{|\alpha|^{2n}}{n!},
\end{equation}
that is a Poisson distribution with:
$$ \langle \hn \rangle = |\alpha|^2 \quad \mbox{and}\quad \Delta^2(\hn) = |\alpha|^2,$$
as the reader can easily check. $ $ \hfill $\blacksquare$
\end{example}

If we calculate the expectation of the quadrature operator we find:
\begin{equation}
\langle \hx_\theta \rangle =
\bra{\alpha} \ha \ket{\alpha}\, \e^{-i\theta} + 
\bra{\alpha} \ha^{\dag} \ket{\alpha}\, \e^{i\theta} =
2\, \Real\left[\alpha\, \e^{-i\theta}\right],
\end{equation}
and if we put $\alpha = (x_1 + i x_2)/2$, with $x_1,x_2 \in {\mathbbm R}$, we obtain:
\begin{equation}
\langle \hx_\theta \rangle = x_1\, \cos\theta + x_2\, \sin\theta,
\end{equation}
as for a classical wave, see Eq.~(\ref{x:theta:class}). However, whereas in the classical case the uncertainty of the expectation of the quadrature is null, in the present case we have:
\begin{equation}
\Delta^2( \hx_\theta) = \langle \hx_\theta^2 \rangle - \langle \hx_\theta \rangle^2 = 1,\quad 
\forall \theta,
\end{equation}
and, in particular:
\begin{equation}
\Delta^2(\hx_1)\,\Delta^2(\hx_2) = 1.
\end{equation}
Since, in general, given two operators $\hat{A}$ and $\hat{B}$ we have:
\begin{equation}
\Delta^2(\hat{A})\,\Delta^2(\hat{B}) \ge
\frac14 \left| \left\langle \left[ \hat{A},\hat{B} \right] \right\rangle \right|^2 ,
\end{equation} 
and, in our case, $[\hx_1,\hx_2] = 2 i$, we have found that coherent states are \emph{minimun uncertainty states}.
\par
Two coherent states $\ket{\alpha}$ and $\ket{\beta}$ are not orthogonal:
\begin{align}
\braket{\alpha}{\beta} &= \exp\left(-\frac{|\alpha|^2 + |\beta|^2}{2}\right)
\sum_{n=0}^{\infty} \frac{(\alpha^*)^n \beta^n}{n!}\\
&= \exp\left(-\frac{|\alpha-\beta|^2}{2}\right)
\, \exp\left(\frac{\alpha^*\beta - \alpha\beta^*}{2}\right),
\end{align}
and we have:
\begin{equation}
|\braket{\alpha}{\beta}|^2 = \e^{-|\alpha-\beta|^2} \ne 0.
\end{equation}
On the other hand, coherent states are \emph{overcomplete}, in fact we have the following resolution of the identity:
\begin{equation}\label{complete:ch:st}
\frac{1}{\pi} \int_{\mathbbm C} \ket{\alpha}\bra{\alpha}\, \d^2\alpha = \idop.
\end{equation}

\begin{exercise}
Prove Eq.~(\ref{complete:ch:st}). (Hint: use the completeness relation of the Fock states and the polar representation of complex numbers$\ldots$) $ $ \hfill $\blacksquare$
\end{exercise}

\subsection{Thermal states}
Up to now we have met two \emph{pure} states, namely, the photon number states $\ket{n}$ and the coherent states $\ket{\alpha}$. Another common state is the so-called thermal state, which is a \emph{mixed} state, being a mixture of Fock states.
The thermal state of the single-mode free-field Hamiltonian can be written as:
\begin{subequations}\label{th:s}
\begin{align}
\hrho_{\rmpl{th}}(N_{\rmpl{th}}) &= \frac{1}{1+N_{\rmpl{th}}}
 \left( \frac{N_{\rmpl{th}}}{1+N_{\rmpl{th}}} \right)^{\ha^{\dag} \ha},\\[1ex]
 &= \frac{1}{1+N_{\rmpl{th}}}
\sum_{n=0}^{\infty} \left( \frac{N_{\rmpl{th}}}{1+N_{\rmpl{th}}} \right)^n \ket{n}\bra{n}.
\end{align}
\end{subequations}
where $N_{\rmpl{th}}$ represents the average number of photons. It is straightforward to verify that the purity $\mbox{Tr}[\hrho_{\rmpl{th}}^2]<1$ if $N_{\rmpl{th}} >0$; if $N_{\rmpl{th}} = 0$ we have $\hrho_{\rmpl{th}}(0) = \ket{0}\bra{0}$, that is the vacuum state. Of course, the photon number statistics is given by:
\begin{equation}
p(n) = \frac{1}{1+N_{\rmpl{th}}} \left( \frac{N_{\rmpl{th}}}{1+N_{\rmpl{th}}} \right)^n.
\end{equation}
Note that, in particular, the state of radiation at thermal equilibrium is obtained by Eqs.~(\ref{th:s}), where now the average number of photons reads \cite{planck}:
\begin{equation}
N_{\rmpl{th}}(\omega, T)  = \frac{1}{\exp[\hbar\omega/(k_{\rmpl{B}}T)]-1},
\end{equation}
$\omega$ being the radiation frequency, $T$  the temperature and $k_{\rmpl{B}}$ the Boltzmann constant. This explains why the states described by Eqs.~(\ref{th:s}) are usually called ``thermal states''.

\begin{exercise}
Show that for a thermal state $\hrho_{\rmpl{th}}(N_{\rmpl{th}})$ one has $\langle \hn \rangle = \mbox{Tr}[\hrho_{\rmpl{th}}\, \hn] = N_{\rmpl{th}}$ and $\Delta^2(\hn) = N_{\rmpl{th}}(N_{\rmpl{th}} + 1)$.
$ $ \hfill $\blacksquare$
\end{exercise}

\begin{exercise}
Show that for a thermal state $\hrho_{\rmpl{th}}(N_{\rmpl{th}})$ one obtains $\langle \hx_\theta \rangle = 0$ and variance $\Delta^2(\hx_\theta) = 2 N_{\rmpl{th}} + 1$. $ $ \hfill $\blacksquare$
\end{exercise}

\section{Function of operators and ordering theorems}\label{s:theorems}
Given the Hermitian operator $\hA$, such that $\hA \ket{\psi_n} = A_n \ket{\psi_n}$, and a function $f(x)$ with Maclaurin expansion:
\begin{equation}
f(x) = \sum_{l=0}^{\infty}\frac{1}{l!}\, f^{(l)}(0)\, x^l,
\end{equation}
it follows:
\begin{equation}
f\left(\hA\right) \ket{\psi_n} = f(A_n) \ket{\psi_n} \Rightarrow
f\left(\hA\right) = \sum_n f(A_n) \ket{\psi_n}\bra{\psi_n}.
\end{equation}

\begin{example}
A unitary operator $\hU$ can be always written as $\hU = \exp\left(i \hB\right)$, where $\hB$ is Hermitian. Therefore we have:
$$ \hU = \sum_{k=0}^{\infty} \frac{1}{k!}\, \left( i \hB \right)^k.$$
$ $ \hfill $\blacksquare$
\end{example}

\par
Given two operators $\hA$ and $\hB$ we have the following theorems \cite{barnett}:

\begin{thm}\label{thm:exp:sum}
If $\left[\hA,\hB\right] \in {\mathbbm C}$, we have:
\begin{subequations}
\begin{align}
e^{\hA+\hB}&= e^{\hA}\, e^{\hB}\, e^{-\frac12 \left[\hA,\hB\right]},\\[1ex]
&=
e^{\hB}\, e^{\hA}\, e^{\frac12 \left[\hA,\hB\right]}\,.
\end{align}
\end{subequations}
$ $ \hfill $\blacksquare$
\end{thm}

\begin{thm}\label{thm:evol}
\begin{align}
\e^{\hA}\, \hB\, \e^{-\hA} = \hB + &\left[\hA,\hB\right] + 
\frac{1}{2!} \left[\hA,\left[\hA,\hB\right] \right] \nonumber\\[1ex]
&+ \frac{1}{3!} \left[\hA,\left[\hA,\left[\hA,\hB\right]\right] \right] + \cdots
\label{thm:evol:Eq}
\end{align}
$ $ \hfill $\blacksquare$
\end{thm}

Theorem~\ref{thm:exp:sum} can be used in the case of Schr\"{o}dinger evolution and in the presence of a Hamiltonian, which can be reduced to the form $\hH \propto \hA+\hB$ where $\left[\hA,\hB\right] \in {\mathbbm C}$. On the other hand, Theorem~\ref{thm:evol} is extremely useful in the case of the Heisenberg evolution of the operators under the action of a Hamiltonian $\hH$.  In this case $\hA = -i \hH t / \hbar$ and $\hB$ is the operator under investigation.

\begin{example}{\bf (The displacement operator)} The so-called displacement operator is:
\begin{equation}\label{displacement}
\hD(\alpha) = \exp\left( \alpha \ha^{\dag} - \alpha^* \ha \right).
\end{equation}
If we apply $\hD(\alpha)$ to the vacuum state and use Theorem~\ref{thm:exp:sum} we have:
\begin{align*}
\hD(\alpha)\ket{0} &= \e^{-|\alpha|^2/2}\exp(\alpha\ha^{\dag})
\underbrace{\exp(\alpha^*\ha)\ket{0}}_{\displaystyle \ket{0}}\\
&= \e^{-|\alpha|^2/2}
\sum_{n=0}^{\infty} \frac{1}{n!}\left(\alpha \ha^{\dag}\right)^n
\ket{0}\\
&= \e^{-|\alpha|^2/2}
\sum_{n=0}^{\infty} \frac{\alpha^n}{\sqrt{n!}}\ket{n},
\end{align*}
that is $\hD(\alpha)\ket{0} = \ket{\alpha}$: we have ``displaced'' the vacuum state and obtained a coherent state. We can associate the linear Hamiltonian $\hH = i\hbar(g \ha^{\dag}-g^* \ha)$
with the displacement operator. In fact, we find the following evolution operator:
$$
\exp\left(- i \frac{\hH}{\hbar} t\right) = \hD(g t),
$$
that is a displacement with amplitude $\alpha = g t$.
$ $ \hfill $\blacksquare$
\end{example}

\begin{example}
In this example we evaluate the Heisenberg evolution of the annihilation operator under the action of the displacement operator. We can apply the Theorem~\ref{thm:evol} with $\hA = \alpha^* \ha - \alpha \ha^{\dag}$ and $\hB = \ha$, since  $\left[\hA,\hB \right] = \alpha$ only the first two terms of the r.h.s.~of Eq.~(\ref{thm:evol:Eq}) survive, namely:
\begin{equation}\label{D:mode:evol}
\hD^{\dag}(\alpha)\,\ha\, \hD(\alpha) = \ha + \alpha.
\end{equation}
$ $ \hfill $\blacksquare$
\end{example}

\begin{exercise}
Calculate $\hD^{\dag}(\alpha)\,\hx_\theta \, \hD(\alpha)$ and, using the result, calculate $\bra{\alpha} \hx_{\theta} \ket{\alpha}$, where $\ket{\alpha}$ is a coherent state.
$ $ \hfill $\blacksquare$
\end{exercise}

\begin{figure}[t]
\begin{center}
    \includegraphics[width=0.4\columnwidth]{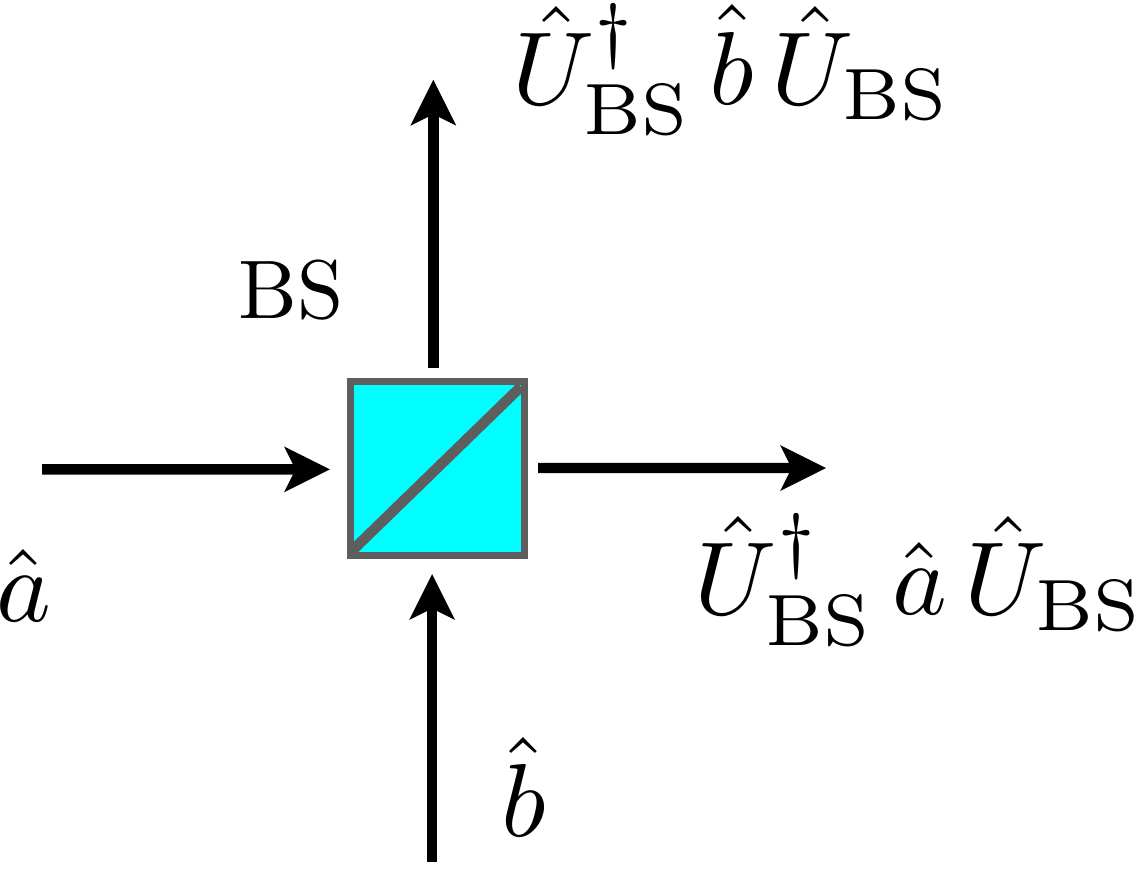}
\caption{\label{f:BS} Scheme of a beam splitter.}
\end{center}
\end{figure}
\section{The beam splitter}\label{s:BS}
The beam splitter (BS) is one of the most common devices we can find in any quantum optics experiment \cite{GK}. The schematic representation of a BS is given in Fig.~\ref{f:BS}. The Hamiltonian describing the interaction through a BS involves two input modes, $\ha$ and $\hb$, namely, $\hH \propto \ha^{\dag}\hb + \ha\hb^{\dag}$, and the corresponding evolution operator can be written as:
\begin{equation}\label{evol:BS}
\Ubs (\zeta)= \exp\left( \zeta\, \ha^{\dag}\hb - \zeta^{*}\,\ha\hb^{\dag} \right),
\end{equation}
where $[\ha,\hb]=0$ and $\zeta= \phi\, \e^{i\theta}$. Since the two-boson operators $\hJ_{+} = \ha^{\dag}	\hb$, $\hJ_{-} = \ha\hb^{\dag}$ and $\hJ_{3} = \frac{1}{2}[\hJ_{+},\hJ_{-}] = \frac{1}{2}(\ha^{\dag}\ha - \hb\hb^{\dag})$ are a realization of the $\mbox{SU}(2)$ algebra, we can rewrite Eq.~(\ref{evol:BS}) as follow:
\begin{subequations}
\begin{align}
\Ubs (\zeta) &=
\exp\left[\e^{i\theta}\tan\phi\, \ha^{\dag}\hb \right]\,
\left(\cos^2\phi\right)^{-(\ha^{\dag}\ha - \hb^{\dag}\hb)/2}\,\nonumber\\
&\hspace{2cm} \times \exp\left[-\e^{-i\theta}\tan\phi\, \ha\hb^{\dag} \right] , \\[1ex]
&=
\exp\left[-\e^{-i\theta}\tan\phi\, \ha\hb^{\dag} \right]\,
\left(\cos^2\phi\right)^{(\ha^{\dag}\ha - \hb^{\dag}\hb)/2}\, \nonumber\\
&\hspace{2cm} \times \exp\left[\e^{i\theta}\tan\phi\, \ha^{\dag}\hb \right] ,
\end{align}
\end{subequations}
which can be useful to calculate the Schr\"{o}dinger evolution of two states through a BS.
Using the Theorem~\ref{thm:evol} it is easy to show that:
\begin{subequations}\label{heisenberg:BS}
\begin{align}\label{BS:heisenberg}
\Ubs^{\dag}(\zeta)\, \ha\, \Ubs(\zeta) &= \ha \cos\phi + \hb\, \e^{i\theta} \sin \phi ,
\\[1ex]
\Ubs^{\dag}(\zeta)\, \hb\, \Ubs(\zeta) &= \hb \cos\phi - \ha\, \e^{-i\theta} \sin \phi ,
\end{align}
\end{subequations}
that can be used to evaluate the Heisenberg evolution of some functions of the field operators. Due to the action of a BS  on the two input modes, this kind of interaction is also called \emph{two-mode mixing interaction}.

\begin{exercise}
The state of a laser beam can be well approximated with a coherent state $\ket{\alpha} = \hD(\alpha)\ket{0}$ of its field mode $\ha$, $\alpha \in {\mathbbm R}$. Assuming that the mode $\hb$ is initially in its vacuum state, calculate the evolution of the input state $\ket{\alpha}\otimes\ket{0}$ through a balanced BS, that is a BS with $\phi=\pi/4$ and $\theta=0$. Do the same calculation with mode $\hb$ excited in a coherent state $\ket{\beta}$, $\beta = |\beta|\, \e^{i\varphi}$. (Hint: consider the Heisenberg evolution of the displacement operators.)
$ $ \hfill $\blacksquare$
\end{exercise}

\begin{example}
{\bf (Hong-Ou-Mandel effect)} Let's assume that two single photons interacts through a balanced BS ($\phi=\pi/4$ and $\theta=0$): what is the two-mode output state? We can write the two-mode input state of mode $\ha$ and $\hb$ as: $$\ket{1}\ket{1} = \ket{1}\otimes \ket{1} = \ha^{\dag}\hb^{\dag}\ket{0}.$$ The output state is thus given by $\Ubs \ket{1}\ket{1}$, with $\Ubs = \Ubs (\frac{\pi}{4})$. Exploiting Eqs.~(\ref{heisenberg:BS}) we have:
\begin{align*}
\Ubs\ket{1}\ket{1} &= \Ubs \, \ha^{\dag}\hb^{\dag}\, \Ubs^{\dag}
\underbrace{\Ubs \ket{0}}_{\displaystyle \ket{0}}, \\[1ex]
&= \left( \frac{\ha^{\dag} - \hb^{\dag}}{\sqrt{2}} \right)
\left( \frac{\hb^{\dag} + \ha^{\dag}}{\sqrt{2}} \right) \ket{0} = \frac{\ket{2}\ket{0} - \ket{0}\ket{2}}{\sqrt{2}}.
\end{align*}
Therefore if we put two photodetectors at the BS outputs, we never obtain coincidence counts that correspond to the output state $\ket{1}\ket{1}$.  $ $ \hfill $\blacksquare$
\end{example}

\section{Single-mode squeezing}\label{s:smSq}
When the value of a quadrature variance is less than the vacuum state one, in our case less than 1, we say that the state is ``squeezed''. Squeezing transformations correspond to Hamiltonians
of the form $\hH \propto (\ha^{\dag})^2 + \ha^2$ and the evolution operator can be written as \cite{MW}:
\begin{equation}\label{sq:op}
\hS(\xi) = \exp\left[\mbox{$\frac12$} \xi (\ha^{\dag})^2 -
\mbox{$\frac12$} \xi^{*} \ha^2 \right],
\end{equation}
where $\xi = r\, \e^{i\psi}$. If we define the operators $\hK_{+} = \frac12 (\ha^{\dag})^2 $, 
$\hK_{-} = \frac12 \ha^2 $ and $\hK_3 = - \frac12 [\hK_{+} ,\hK_{-} ] = \frac12 (\ha^{\dag}\ha + \frac12)$, we obtain a boson realization of $\mbox{SU}(1,1)$ algebra. In particular, we can write:
\begin{equation}\label{sq:s11}
\hS(\xi) = \exp\left[ \frac{\nu}{2\mu}(\ha^{\dag})^2\right]\,
\mu^{-(\ha^{\dag}\ha + \frac12)}\,
\exp\left[ -\frac{\nu^{*}}{2\mu}\ha^2\right],
\end{equation}
where $\mu = \cosh r$ and $\nu = \e^{i\psi} \sinh r$.
The evolution of the annihilation operator under the action of the squeezing operator (\ref{sq:op}) reads:
\begin{equation}\label{sq:mode:evol}
\hS^{\dag}(\xi)\,\ha\,\hS(\xi) = \mu \ha + \nu \ha^{\dag}.
\end{equation}

\begin{example}\label{squeezed:pn}
{\bf (Squeezed vacuum)} If we apply the squeezing operator to the vacuum state, we obtain the squeezed vacuum:
$$\hS(\xi)\ket{0} = \frac{1}{\sqrt{\mu}} \sum_{n=0}^{\infty} \left(\frac{\nu}{2\mu}\right)^n
\frac{\sqrt{(2n)!}}{n!} \, \ket{2n},$$
which can be easily calculated using Eq.~(\ref{sq:s11}) and the expansion of the exponential. It is worth noting that the squeezed vacuum is a superposition of only even number states.
$ $ \hfill $\blacksquare$
\end{example}

\begin{exercise}\label{ex:sq:parameters}
Let us assume that $\xi = r \in {\mathbbm R}$. Show that:
\begin{align*}
(i)&\quad  \langle \hn \rangle = \sinh^2 r; \\
(ii)&\quad \Delta^2(\hn) = 2 \sinh^2 r (\sinh^2 r + 1);\\
(iii)&\quad  \langle \hx_{\theta} \rangle = 0, \quad \forall \theta; \\
(iv)&\quad  \Delta^2( \hx_{\theta}) = \e^{2r}\cos^2\theta + \e^{-2r}\sin^2\theta.
\end{align*}
$ $ \hfill $\blacksquare$
\end{exercise}

Starting from the results of the Exercise~\ref{ex:sq:parameters} and focusing on the quadrature operators $\hx_1$ and $\hx_2$, obtained from $\hx_\theta$ with $\theta = 0$ and $\theta = \pi/2$, respectively, we have:
\begin{equation}
\Delta^2(\hx_1) = \e^{2r} \quad \mbox{and} \quad
\Delta^2(\hx_2) = \e^{-2r},
\end{equation}
namely, the variance of the quadrature $\hx_2$ is below the vacuum level if $r>0$. It is also worth noting that $\Delta^2(\hx_1) \, \Delta^2(\hx_2) = 1$ that is the squeezed vacuum is a minimum uncertainty state, but with different quadrature variances.

\begin{exercise}
Prove that the displaced squeezed state $\hD(\alpha)\hS(\xi)\ket{0}$ is still a minimum uncertainty state. $ $ \hfill $\blacksquare$
\end{exercise}

\section{Two-mode squeezing}\label{s:tmSq}
The two-mode counterpart of the single-mode squeezing corresponds to Hamiltonians of the form $\hH \propto \ha^{\dag}\hb^{\dag} + \ha\hb$ and the evolution operator reads  \cite{MW}:
\begin{equation}\label{sq:2:op}
\hS_2(\xi) = \exp\left(\xi \ha^{\dag}\hb^{\dag} -\xi^{*} \ha\hb \right),
\end{equation}
where $\xi = r\, \e^{i\psi}$. If we define the operators $\hK_{+} = \ha^{\dag} \hb^{\dag}$, 
$\hK_{-} = \ha \hb$ and $\hK_3 = -\frac12 [\hK_{+} ,\hK_{-} ] = \frac12 (\ha^{\dag}\ha + \hb^{\dag}\hb + 1)$, we obtain a two-boson realization of $\mbox{SU}(1,1)$ algebra. As in the case of single-mode squeezing, we can write:
\begin{equation}\label{sq:2:s11}
\hS_2(\xi) = \exp\left( \frac{\nu}{\mu}\ha^{\dag}\hb^{\dag}\right)\,
\mu^{-(\ha^{\dag}\ha + \hb^{\dag}\hb + 1)}\,
\exp\left( -\frac{\nu^{*}}{\mu}\ha\hb\right),
\end{equation}
where $\mu = \cosh r$ and $\nu = \e^{i\psi} \sinh r$. 
The evolution of $\ha$ and $\hb$ under the action of $\hS_2(\xi)$ leads to:
\begin{subequations}\label{two:sq:mode:evol}
\begin{align}
\hS^{\dag}_2(\xi)\,\ha\,\hS_2(\xi) &= \mu \ha + \nu \hb^{\dag}\,, \\[1ex]
\hS^{\dag}_2(\xi)\,\hb\,\hS_2(\xi) &= \mu \hb + \nu^* \ha^{\dag}.
\end{align}
\end{subequations}

\begin{example}
{\bf (Two-mode squeezed vacuum or twin-beam state)} The reader can easily check that:
\begin{equation}\label{TWB}
\hS_2(\xi)\ket{0} = \frac{1}{\sqrt{\mu}} \sum_{n=0}^{\infty} \left(\frac{\nu}{\mu}\right)^n
 \, \ket{n}\ket{n},
\end{equation}
or, if we introduce the parameter $\lambda = \e^{i\psi} \tanh r$:
\begin{equation}
\hS_2(\xi)\ket{0} = \sqrt{1-|\lambda|^2} \sum_{n=0}^{\infty} \lambda^n
 \, \ket{n}\ket{n},
\end{equation}
that is the two-mode squeezed vacuum or twin-beam state, since a measurement of the photon number on the two beams always leads to the same result.  If we introduce $N = \sinh^2 |\xi|$
the twin-beam state mean number of photons is $\langle \hn \rangle = 2 N$ and $$|\lambda|^2 = \frac{N}{N+1}.$$
$ $ \hfill $\blacksquare$
\end{example}

It is worth noting that the twin-beam state (\ref{TWB}) is a continuous-variable \emph{maximally entangled state}. In fact, in the presence of a pure states of two subsystems $\hrho_{\rmpl{AB}}$, entanglement can be quantified by the excess von Neumann entropy, namely, $\cE(\hrho_{\rmpl{AB}}) = \cS(\hrho_{\rmpl{A}}) + \cS(\hrho_{\rmpl{A}}) - \cS(\hrho_{\rmpl{AB}})$, where $\hrho_{\rmpl{A}} = \mbox{Tr}_{\rmpl{B}}[\hrho_{\rmpl{AB}}]$ and $\hrho_{\rmpl{B}}= \mbox{Tr}_{\rmpl{A}}[\hrho_{\rmpl{AB}}]$ are the density operators of the two subsystems and $\cS(\hrho) = - \mbox{Tr}[\hrho \log \hrho]$ is the von Neumann entropy. In the case of the twin-beam state (\ref{TWB}), the two subsystems are described by two thermal states with the same mean photon number $N = \sinh^2|\xi|$. Since a thermal state maximizes the von Neumann entropy for a fixed energy $N$, the excess von Neumann entropy reaches its maximum: the twin-beam state is maximally entangled.

\begin{exercise}
Given the twin-beam state $\hS_2(\xi)\ket{0}$, explicitly write the expression of the density operators of the two subsystems and calculate the excess von Neumann entropy. $ $ \hfill $\blacksquare$
\end{exercise}

\begin{exercise}
Given the two squeezed states $\hS_{\ha}(-r)\ket{0}$ and $\hS_{\hb}(r)\ket{0}$ of modes $\ha$ and $\hb$, respectively, calculate the two-mode state obtained after their interference through a  balanced BS. Assume $r \in {\mathbbm R}$ and $r >0$. $ $ \hfill $\blacksquare$
\end{exercise}

\section{Photon number statistics and moment generating function}\label{s:pn}
In this section we address a useful method for studying the photon number statistics $p(n) = \bra{n}\hrho\ket{n}$ of a state $\hrho$. The moment generating function is defined as \cite{barnett}:
\begin{equation}
M(\mu) = \sum_{n=0}^{\infty} (1-\mu)^n p(n),\quad (0 \le \mu \le 2).
\end{equation}
Note that $M(0) = \sum_{n=0}^{\infty} p(n) = 1$, whereas:
\begin{equation}
\frac{(-1)^n}{n!} \frac{d^n}{d\mu^n}\, M(\mu)\Big|_{\mu=1} = p(n).
\end{equation}
We also note that $M(2) = \sum_{n} p(2 n) - \sum_{n} p(2 n +1)$ corresponds to the difference between the probability that the photon number is even and the probability that it is odd.
Furthermore, given $M(\mu)$ we can calculate the $m$-th moment of $p(n)$ as:
\begin{equation}
\left[ (\mu - 1) \frac{d}{d\mu} \right]^m M(\mu)\Big|_{\mu=0} = 
\sum_{n=0}^{\infty} n^m p(n) \equiv \langle \hn^m \rangle.
\end{equation}
\par
In the case of a coherent state $\ket{\alpha}$ we obtain the following expression for the moment generating function:
\begin{equation}\label{mgf:coherent}
\ket{\alpha} \Rightarrow M(\mu) = 
\e^{-|\alpha|^2}\sum_{n=0}^{\infty} (1-\mu)^n \frac{|\alpha|^{2n}}{n!} = \e^{-\mu |\alpha|^2}.
\end{equation}
In the presence of a thermal state $\hrho_{\rmpl{th}}(N_{\rmpl{th}})$ we have:
\begin{align}
M(\mu) &=
\frac{1}{1+N_{\rmpl{th}}} \sum_{n=0}^{\infty} (1-\mu)^n
\left(\frac{N_{\rmpl{th}}}{1+N_{\rmpl{th}}}\right)^n \\[1ex]
&= \frac{1}{1+ \mu N_{\rmpl{th}}}.
\label{mgf:thermal}
\end{align}
It is interesting to note that in both the cases of the coherent state and of the thermal state, the parameter $\mu$ of the moment generating function multiplies the mean energy, that is $|\alpha|^2$ or $N_{\rmpl{th}}$, respectively. This is not the case if we consider, for example, a Fock state:
\begin{equation}\label{mgf:fock}
\ket{n}\quad\Rightarrow\quad M(\mu) = (1-\mu)^n,
\end{equation}
or the squeezed vacuum state:
\begin{equation}\label{mgf:squeezed}
\hS(\xi)\ket{0} \Rightarrow M(\mu) = \frac{1}{\sqrt{1+2\mu N - \mu^2 N}},
\end{equation}
with $N=\sinh^2 |\xi|$. In this last case we can easily see that $M(2) = 1$: this directly follows from the photon number statistics of the squeezed vacuum that involves only the even terms (see Example~\ref{squeezed:pn}).

\begin{figure}[t]
\begin{center}
    \includegraphics[width=0.45\columnwidth]{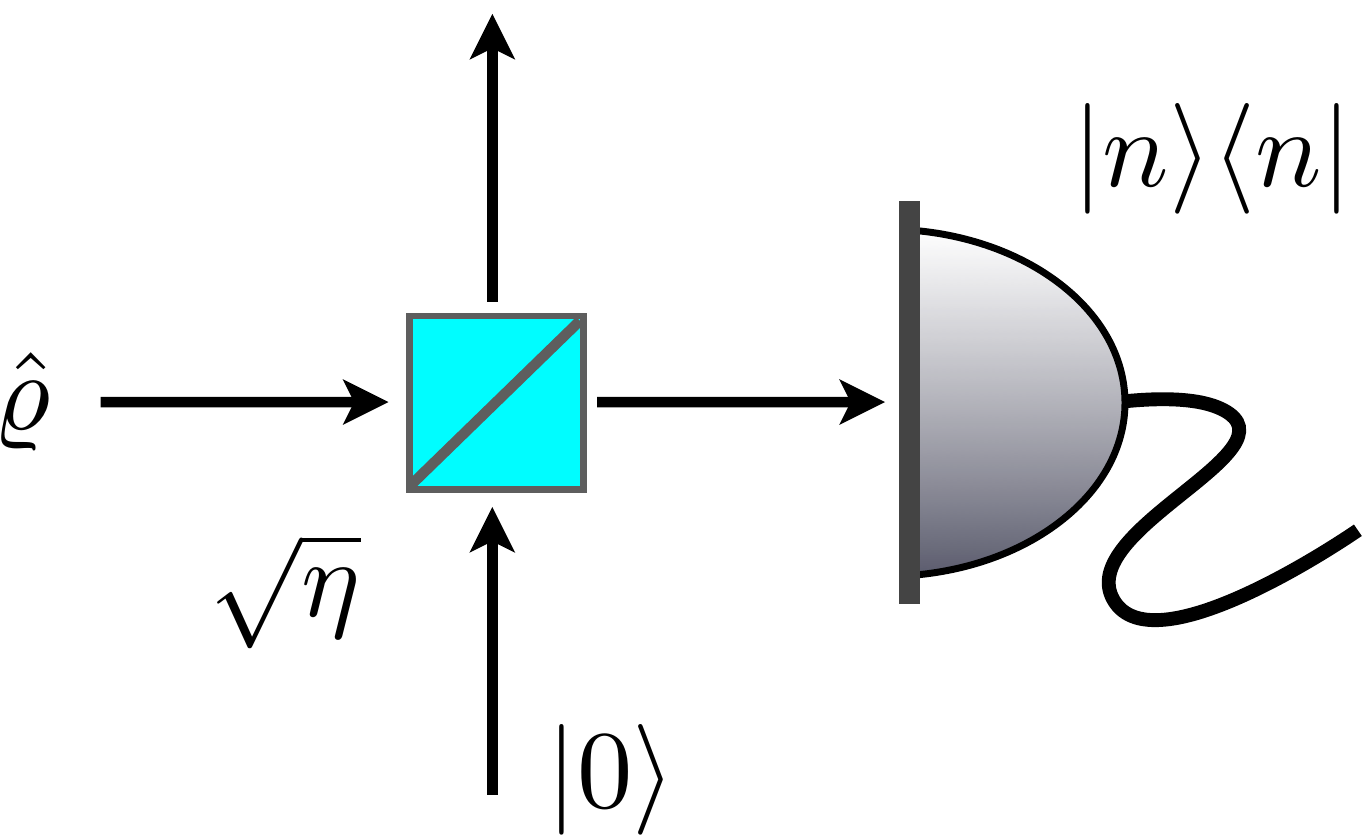}
\caption{\label{f:detection} A photon number resolving detector with quantum efficiency $\eta$ can be represented as an ideal photodetector and a BS with transmissivity $\eta$ in front of it.}
\end{center}
\end{figure}
\subsection{Bernoulli sampling from non-unit efficiency photodetection} A photon number resolving detector allows to directly measure the photon number distribution $p(n) = \bra{n} \hrho\ket{n}$ of an input state $\hrho$ and is described by the projectors $\ket{n}\bra{n}$ onto the photon number basis. However, a realistic detector has a non-unit quantum efficiency $\eta$, that can be seen as an overall loss of photons during the detection process. From the theoretical point of view, a real photodetector can be modeled as a BS with transmissivity $\eta$ and an ideal photon number resolving detector, as sketched in Fig.~\ref{f:detection}. As a matter of fact, if we send a single photon state $\ket{1}$ to the realistic detector, $\eta$ corresponds to the probability to detect it. What happens when we send a Fock state $\ket{l}$, with $l>1$? Let us focus on Fig.~\ref{f:detection}. The two-mode state just after the BS is:
\begin{align}
\Ubs &\ket{l}\ket{0} = \frac{1}{\sqrt{l!}} \left(\sqrt{\eta}\, \ha^{\dag} -
\sqrt{1-\eta}\, \hb^{\dag}\right)^l \ket{0}\notag\\
&= \frac{1}{\sqrt{l!}} \sum_{k=0}^{l}{l \choose k} (-1)^{k} \sqrt{\eta^{l-k} (1-\eta)^{k}} \,
(\ha^{\dag})^{l-k}(\hb^{\dag})^k \ket{0}\notag\\
&= \sum_{k=0}^{l}(-1)^{k} \sqrt{{l \choose k} \eta^{l-k} (1-\eta)^{k}} \,
\ket{l-k}\ket{k}\notag\\
&= \sum_{m=0}^{l}(-1)^{l-m} \sqrt{P(m;\eta)} \,
\ket{m}\ket{l-m}.
\end{align}
where
\begin{equation}
P(m;\eta) = {l \choose m} \eta^{m} (1-\eta)^{l-m}
\end{equation}
is the probability to detect $m$ photons, $m\le l$. Of course, if $\eta \to 1$ we have $P(m; \eta) \to p(m) = \delta_{l,m}$.
\par
It is now clear that if we know the \emph{actual} photon number statistics $p(m)$ of the state $\hrho$, then the \emph{detected} photon number statistics is:
\begin{equation}
P(m;\eta) = \sum_{l=0}^{\infty} {l \choose m} \eta^{m} (1-\eta)^{l-m} p(l),
\end{equation}
and the corresponding moment generating function is:
\begin{align}
M(\mu;\eta) &= \sum_{m=0}^{\infty} (1-\mu)^m P(m;\eta) \\
&= \sum_{l=0}^{\infty} (1-\eta \mu)^l p(l) \equiv M(\eta \mu),\label{mgf:eta}
\end{align}
that is the moment generating function of the actual photon distribution but with the substitution $\mu \to \eta\mu$.

\begin{exercise}
Apply the result of Eq.~(\ref{mgf:eta}) to the case of a coherent, thermal and squeezed state. What are the main differences? What is the physical insight? $ $ \hfill $\blacksquare$
\end{exercise}

\begin{figure}[t]
\begin{center}
    \includegraphics[width=0.55\columnwidth]{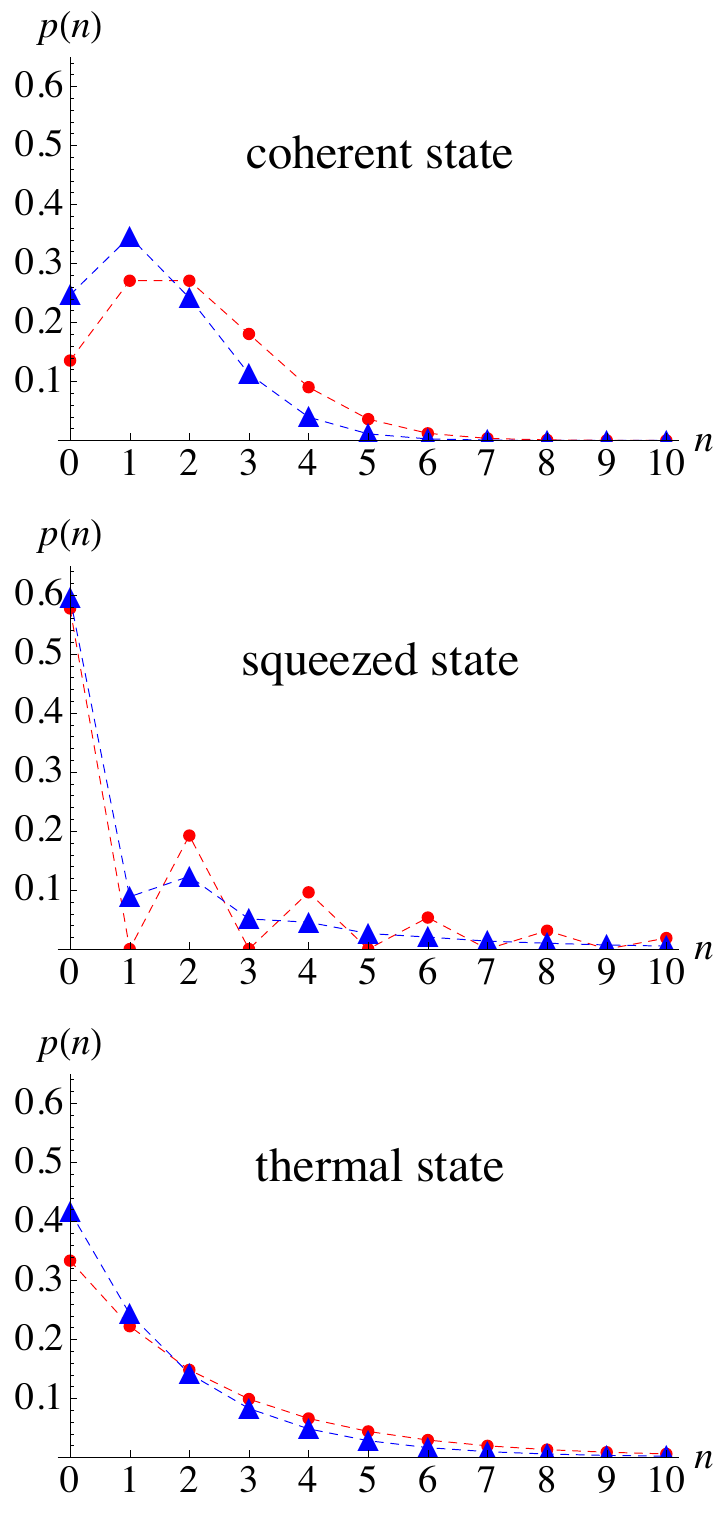}
    \vspace{-0.3cm}
\caption{\label{f:pn} Photon number distribution of a thermal coherent state, thermal state and squeezed vacuum state in the ideal case ($\eta=1$, circles) and in the presence of a non-unit quantum efficiency ($\eta = 0.7$, triangles). In all the cases we set $\langle \hn \rangle = 2.0$.}
\end{center}
\end{figure}
In Fig.~\ref{f:pn} we show the effect of a non-unit quantum efficiency on the photon number distribution of a coherent, a thermal and a squeezed vacuum state.

\section{The Fano factor and the Mandel parameter}\label{s:fano}
In order to classify the photon number distribution of an optical state, it is somehow convenient to use its width relative to a Poisson distribution. The Fano factor is defined as \cite{MW}:
\begin{equation}
\cF = \frac{\Delta^2(\hn)}{\langle \hn \rangle},
\end{equation}
that is just the ratio between the actual width of the photon number distribution and the width of a Poissonian with the same mean photon number. If $\cF >1$ we are in the presence of \emph{super-Poissonian} states; if $\cF < 1$ we have \emph{sub-Poissonian} states. Light with Poissonian ($\cF=0$) or super-Poissonian properties can be also described within a semiclassical theory, where the field is, in general, a classical superposition of electromagnetic waves, but the detector is a quantum device (for example an atom). Within the semiclassical framework, a Poissonian light can be modelled as the classical wave in Eq.~(\ref{class:wave}), whereas a super-Poissonian one is obtained by the superposition of a large number of uncorrelated waves with the same amplitude and frequency, but random phases. However, it is not possible to give an account of sub-Poissonian light within the semiclassical framework \cite{GK}, which turns out to be nonclassical and, thus, requires a quantum description of light. It is worth noting, that the sub-Poissoninan character of the light is only a {\it sufficient} condition for nonclassicality (squeezed states exhibit a super-Poissonian photon statistics but are nonclassical, as we shall see).

A typical example of sub-Poissonian and, in turn, nonclassical states is given by the Fock states, for which we find $\cF=0$. A coherent state has $\cF=1$. The thermal state with $N_{\rmpl{th}}$ mean photons is super-Poissonian, since $\cF = N_{\rmpl{th}} + 1$. Also squeezed states are super-Poissonian and we have $\cF = 2 (\langle \hn \rangle + 1)$: therefore given a thermal state and a squeezed state with the same mean number of photons, the squeezed state has a Fano factor that is twice the thermal states one.
\par
As a matter of fact, different quantum states may have the same Fano factor: both the coherent state $\ket{\alpha}$ and the mixed state $\hrho = \e^{-|\alpha|^2}\sum_{n} (n!)^{-1} |\alpha|^{2n}\ket{n}\bra{n}$ have the same Fano factor $\cF=1$.

The Fano factor is directly related to another quantity, the Mandel's $\cQ$-parameter, defined as \cite{GK}:
\begin{align}
\cQ &= \cF -1,\\[1ex]
&=  \langle \hn \rangle \left[ g^{(2)}(0) - 1 \right],\label{g2}
\end{align}
where
\begin{equation}\label{g2}
g^{(2)}(0) = \frac{\langle \hn^2 \rangle - \langle \hn \rangle}{\langle \hn \rangle^2}
\end{equation}
is the normalized second-order correlation function introduced by Glauber \cite{RJG}. We have $\cQ = 0$ and $\cQ > 0$ for Poissoninan and super-Poissonian light, and $-1 < \cQ < 0$ for the sub-Poissonian one. Thanks to Eq.~(\ref{g2}), it is also clear that a sufficient (but not necessary!) condition for nonclassicality is $g^{(2)}(0) < 1$.

\section{Operator ordering and number operator}\label{s:ordering}
We have three possible creation and annihilation orderings:
\begin{itemize}
\item \emph{normal ordering} (creation operators on the left):
$\left(\ha^\dag\right)^n\ha^m$;
\item \emph{antinormal ordering} (creation operators on the right):
$\ha^m\left(\ha^\dag\right)^n$;
\item \emph{symmetric ordering} (balanced sum of all the possible combinations): for example, $[\ha^{\dag}\ha]_{\rmpl{s}} = \frac12 (\ha^{\dag}\ha+\ha\ha^{\dag}).$
\end{itemize}
Depending on the problem and/or the system under investigation, one ordering could be preferred to the others. Given a function $f(\ha,\ha^{\dag})$ of the annihilation and creation operator, we use the notation $:\!\! f(\ha,\ha^{\dag})\!\! :$ for the normal ordering, $\threedotsbin f(\ha,\ha^{\dag})\threedotsbin$ for the antinormal ordering and $[f(\ha,\ha^{\dag})]_{\rmpl{s}}$ for the symmetric ordering. Is this section we focus only on the number operator $\hn$, thus we have:
\begin{itemize}
\item $:\!\! \hn \!\! : = \ha^{\dag}\ha$;
\item $\threedotsbin \hn \threedotsbin = \ha\ha^{\dag}$;
\item $[\ha^{\dag}\ha]_{\rmpl{s}} = \frac12 (\ha^{\dag}\ha+\ha\ha^{\dag}) = \ha^{\dag}\ha + \frac12$.
\end{itemize}

It is straightforward to show that the normalized second-order correlation function (\ref{g2}) can be written as:
\begin{equation}
g^{(2)}(0) = \frac{\langle :\! \hn^2 \!\! : \rangle}{\langle \hn \rangle^2} =
\frac{\langle \ha^{\dag}\ha^{\dag} \ha \ha \rangle}{\langle \ha^{\dag} \ha \rangle^2}.
\end{equation}
Here, the normal ordering of the field operators directly follows from the photodetection theory, where the detector actually absorbs photons \cite{GK,RJG}.

\begin{example}
In this example we first consider a coherent state $\ket{\alpha}$. We find:
$$\bra{\alpha} :\! \hn \! : \ket{\alpha}  = |\alpha|^2,\quad
\bra{\alpha} \threedotsbin \! \hn \! \threedotsbin \ket{\alpha}  = |\alpha|^2 + 1,
$$
and
$$
\bra{\alpha} [ \hn ]_{\rmpl{s}} \ket{\alpha}  = |\alpha|^2 + \mbox{$\frac12$}.
$$
The reader can easily check that:
$$\bra{\alpha} :\!\! \hn^{k} \!\! : \ket{\alpha}  = |\alpha|^{2k},$$ which implies that for a coherent state the normal-ordered variance vanishes, $:\!\!\Delta^2(\hn)\!\! : \, = 0$. From this result follows that:
\begin{itemize}
\item $:\!\!\Delta^2(\hn)\!\! : \, = 0\quad \Rightarrow \quad \mbox{Poissonian}$;
\item $:\!\!\Delta^2(\hn)\!\! : \,  > 0\quad \Rightarrow \quad \mbox{super-Poissonian}$;
\item $:\!\!\Delta^2(\hn)\!\! : \, < 0\quad \Rightarrow \quad \mbox{sub-Poissonian}$.
\end{itemize}
In particular, in the presence of a thermal state $\hrho_{\rmpl{th}}(N_{\rmpl{th}})$ we have $:\!\!\Delta^2(\hn)\!\! : \, = k!\, N_{\rmpl{th}}^k$ and, thus, $:\!\!\Delta^2(\hn)\!\! : \, = N_{\rmpl{th}}^2$.
$ $ \hfill $\blacksquare$
\end{example}

\begin{exercise}
Calculate the normal-, anti-normal- and symmetric-ordered photon number variances in the case of a Fock state $\ket{n}$. $ $ \hfill $\blacksquare$
\end{exercise}

\section{Characteristic functions}\label{s:char}
The use of characteristic functions or quasi-probability distributions allows to obtain a more complete statistical description of the field. They contain all the information necessary to reconstruct the density matrix of the state. The $p$-ordered characteristic function associated with the state $\hrho$ is defined as \cite{barnett}:
\begin{equation}
\chi(\lambda,p) = \mbox{Tr}[\hrho \hD(\lambda)]\, \e^{p |\lambda|^2/2},
\end{equation}
where $\hD(\lambda) = \e^{\lambda \ha^{\dag} - \lambda^* \ha}$ is the displacement operator. We recall that we can rewrite $\hD(\lambda)$ also  in the following three forms:
\begin{subequations}
\begin{align}
\hD(\lambda) &= \e^{\lambda \ha^{\dag}}\, \e^{- \lambda^* \ha} \, \e^{-|\lambda|^2/2},\\
\hD(\lambda) &= \e^{- \lambda^* \ha}\, \e^{\lambda \ha^{\dag}} \, \e^{|\lambda|^2/2},\\
\hD(\lambda) &= \sum_{n,m=0}^{\infty} \frac{\lambda^m (-\lambda^*)^n}{m! n!}\,
\left[(\ha^{\dag})^{m} \ha^n \right]_{\rmpl{s}}.
\end{align}
\end{subequations}
Therefore, according to the value of $p$, we may have the normal ordered characteristic function with $p=1$:
\begin{equation}
\chi(\lambda,1) = \mbox{Tr}[\hrho\, \e^{\lambda \ha^{\dag}}\, \e^{- \lambda^* \ha}], \quad
\mbox{(normal ordered)}
\end{equation}
the antinormal ordered characteristic function with $p=-1$:
\begin{equation}
\chi(\lambda,-1) = \mbox{Tr}[\hrho\, \e^{- \lambda^* \ha}\, \e^{\lambda \ha^{\dag}}], \quad
\mbox{(antinormal ordered)}
\end{equation}
and the symmetric ordered characteristic function with $p=0$:
\begin{equation}
\chi(\lambda,0) = \mbox{Tr}[\hrho\, \e^{\lambda \ha^{\dag} - \lambda^* \ha}], \quad
\mbox{(symmetric ordered).}
\end{equation}
The last form is simply referred to as characteristic function and one writes usually $\chi(\lambda) = \chi(\lambda,0)$.
\par
Starting from $\chi(\lambda,p)$ we can evaluate any $p$-ordered expectation value of a function of $\ha^{\dag}$ and $\ha$, since:
\begin{equation}\label{ordered:moments}
\left(\frac{\partial}{\partial \lambda}\right)^m
\left(-\frac{\partial}{\partial \lambda^*}\right)^n \chi(\lambda,p)\Big|_{\lambda=0} =
\langle (\ha^{\dag})^{m}\, \ha^n\rangle_p, 
\end{equation}
where
\begin{subequations}
\begin{align}
\langle (\ha^{\dag})^{m}\, \ha^n \rangle_1 &= \langle :\!\! (\ha^{\dag})^{m}\, \ha^n \!\!: \rangle,\\
\langle (\ha^{\dag})^{m}\, \ha^n \rangle_{-1} &= \langle \threedotsbin (\ha^{\dag})^{m}\, \ha^n \threedotsbin \rangle,\\[1ex]
\langle (\ha^{\dag})^{m}\, \ha^n \rangle_{0} &= \langle \left[ (\ha^{\dag})^{m}\, \ha^n \right]_{\rmpl{s}} \rangle.
\end{align}
\end{subequations}
\par
We report also the so-called ``Glauber formula'' that allows to connect the density operator $\hrho$ to its characteristic function, namely:
\begin{align}
\hrho &= \frac{1}{\pi} \int_{\mathbbm C} \d^2\lambda\, \chi(\lambda)\, \hD^{\dag}(\lambda)\\[1ex]
&= \frac{1}{\pi} \int_{\mathbbm C} \d^2\lambda\, \mbox{Tr}[\hrho\,\hD(\lambda)]\, \hD^{\dag}(\lambda) \, .
\end{align}

\begin{example}
The $p$-ordered characteristic function of a Fock state $\ket{n}$ is:
$$
\ket{n} \to \chi(\lambda,p) = \e^{(p-1)|\lambda|^2/2}\, L_n(|\lambda|^2)
$$
where: $$L_n(z) = \sum_{m=0}^{n} {n \choose m} \frac{(-z)^m}{m!}$$
are Laguerre polynomials.
\end{example}

\begin{example}
From the previous example follows that the $p$-ordered characteristic function of a thermal state $\hrho_{\rmpl{th}}(N_{\rmpl{th}})$ is given by:
$$
\hrho_{\rmpl{th}} (N_{\rmpl{th}}) \to \chi_{\rmpl{th}}(\lambda,p) =
\exp\left[-\frac12 (1+2 N_{\rmpl{th}}-p) |\lambda|^2 \right]. 
$$
Indeed, if $N_{\rmpl{th}} \to 0$ one has the $p$-ordered characteristic function of the vacuum state, namely:
$$
\ket{0}  \to \chi_{\rmpl{vac}}(\lambda,p) 
= \exp\left[-\frac12 (1-p) |\lambda|^2 \right]. 
$$
It is worth noting that the both $\chi_{\rmpl{th}}(\lambda,p)$ and $\chi_{\rmpl{vac}}(\lambda,p)$ are \emph{Gaussian}: the thermal and the vacuum state belong to the class of the so-called Gaussian states, that are states with Gaussian characteristic (or Wigner, as we will see) functions.
\end{example}

\par
Due to the very definition of the characteristic function, given $\chi(\lambda)$ of the state $\hrho$, the characteristic function $\chi'(\lambda)$  of the state $\hU\hrho\hU^{\dag}$, where $\hU$ is a unitary transformation, is given by:
\begin{align}
\chi'(\lambda) = \mbox{Tr}[\hU\hrho\hU^{\dag}\, \hD(\lambda)] =
\mbox{Tr}[\hrho\,\hU^{\dag} \hD(\lambda) \hU].
\end{align}
In the case of a unitary transformation leading to a linear transformation of the field operators $\ha$ and $\ha^{\dag}$, as in the case of the displacement or squeezing operators, it is straightforward to calculate $\chi'(\lambda)$. In fact, by using Eqs.~(\ref{D:mode:evol}) and (\ref{sq:mode:evol}), respectively, we have:
\begin{align}
\hD(\alpha)\hrho\hD^{\dag}(\alpha) \to \chi'(\lambda) &=
\mbox{Tr}[\hrho\,\hD^{\dag}(\alpha) \hD(\lambda) \hD(\alpha)]
\notag\\
&=\e^{\lambda\alpha^* - \lambda^*\alpha} \,
\mbox{Tr}[\hrho\,\hD(\lambda)]\notag\\
&= \e^{\lambda\alpha^* - \lambda^*\alpha}\, \chi(\lambda).
\end{align}
and (for the sake of simplicity we assume $r\in {\mathbbm R}$):
\begin{align}
\hS(r)\hrho\hS^{\dag}(r) \to \chi'(\lambda) &=
\mbox{Tr}[\hrho\,\hS^{\dag}(r) \hD(\lambda) \hS(r)] \notag\\
&= \mbox{Tr}[\hrho\,\hD(\lambda \cosh r - \lambda^* \sinh r)]
\notag\\
&= \chi(\lambda \cosh r - \lambda^* \sinh r).
\end{align}

\subsection{Trace rule for the characteristic functions} Given two operators $\hA$ and $\hB$ and the corresponding characteristic functions $\chi_{\hA}(\lambda)$ and $\chi_{\hB}(\lambda)$, we have:
\begin{equation}
\mbox{Tr}[\hA\hB] = \frac{1}{\pi} \int_{\mathbbm C} \d^2\lambda\,
\chi_{\hA}(\lambda)\,\chi_{\hB}(-\lambda).
\end{equation}

\section{Quasi-probability distributions}\label{s:QP}
An alternative to the characteristic functions is given by the quasi-probability distributions, which are similar to phase-space distributions \cite{barnett}. A quasi-probability is a real-valued (though it could be negative), normalized function and the moments of products of $\ha^{\dag}$ and $\ha$ are calculated evaluating suitable integrals as in the case of an actual probability distribution.
The $p$-ordered quasi-probability distribution $W(\alpha,p)$ can be defined as the two-dimensional Fourier transform of the corresponding $p$-ordered characteristic function:
\begin{equation}\label{W:p:def}
W(\alpha,p) = \frac{1}{\pi^2} \int_{\mathbbm C}\d^2\lambda\,
\chi(\lambda,p)\, \e^{\alpha\lambda^*-\alpha^*\lambda}.
\end{equation}
We also recall that:
\begin{equation}
\frac{1}{\pi^2} \int_{\mathbbm C}\d^2\lambda\, \e^{\alpha\lambda^*-\alpha^*\lambda}
= \delta^{(2)}(\alpha),
\end{equation}
where $\delta^{(2)}(\alpha)$ is the two-dimensional Dirac's delta function. From the definition 
(\ref{W:p:def}) follows that $W(\alpha,p)$ is normalized:
\begin{align}
\int_{\mathbbm C} \d^2\alpha\, W(\alpha,p) &=  
 \int_{\mathbbm C}\d^2\lambda\, \chi(\lambda,p)
\underbrace{
 \frac{1}{\pi^2} \int_{\mathbbm C}\d^2\alpha\, \e^{\alpha\lambda^*-\alpha^*\lambda}
 }_{\displaystyle \delta^{(2)}(\lambda)}\nonumber
 \\[1ex]
  &= \chi(0,p) = \mbox{Tr}[\hrho]=1.
\end{align}
Furthermore, we have:
\begin{align}
\int_{\mathbbm C}\d^2\alpha\, &W(\alpha,p) \,  (\alpha^*)^{m}\alpha^n\nonumber\\
&=\frac{1}{\pi^2} \int_{\mathbbm C}\d^2\lambda\int_{\mathbbm C}\d^2\alpha\,
\chi(\lambda,p)\, (\alpha^*)^{m}\alpha^n\, \e^{\alpha\lambda^*-\alpha^*\lambda},
\end{align}
but:
$$
(\alpha^*)^{m}\alpha^n\, \e^{\alpha\lambda^*-\alpha^*\lambda} =
\left(-\frac{\partial}{\partial\lambda}\right)^m
\left(\frac{\partial}{\partial\lambda^*}\right)^n
\e^{\alpha\lambda^*-\alpha^*\lambda},
$$
thus, we have:
\begin{align}
\int_{\mathbbm C}\d^2\alpha\, &W(\alpha,p) \,  (\alpha^*)^{m}\alpha^n = \notag\\[1ex]
&=
\frac{1}{\pi^2} \int_{\mathbbm C}\d^2\lambda \,
\chi(\lambda,p)\,\notag\\
&\hspace{1cm} \times \left(-\frac{\partial}{\partial\lambda}\right)^m
\left(\frac{\partial}{\partial\lambda^*}\right)^n
\int_{\mathbbm C}\d^2\alpha\, \e^{\alpha\lambda^*-\alpha^*\lambda},\notag\\[1ex]
&=\frac{1}{\pi^2} \int_{\mathbbm C}\d^2\lambda \,
\chi(\lambda,p)\, \left(-\frac{\partial}{\partial\lambda}\right)^m
\left(\frac{\partial}{\partial\lambda^*}\right)^n
\delta^{(2)}(\lambda)\notag\\[1ex]
&\hspace{0.3cm} \quad \vdots\quad \mbox{(integration by parts)}\quad \vdots \notag\\[1ex]
&= \left(\frac{\partial}{\partial\lambda}\right)^m
\left(-\frac{\partial}{\partial\lambda^*}\right)^n
\chi(\lambda,p) \Big|_{\lambda=0} \notag\\[1ex]
&= \langle (\ha^{\dag})^{m}\, \ha^n\rangle_p
\end{align}
as in the case of Eq.~(\ref{ordered:moments}).
\par
Given $q<p$, we have $\chi(\lambda,q) = \chi(\lambda,p)\,\e^{-(p-q)|\lambda|^2/2}$, and we obtain the following relation between $W(\alpha,q)$ and $W(\alpha,p)$:
\begin{align}\label{W:q:p}
W&(\alpha,q) = \nonumber\\[1ex]
&=
\frac{1}{\pi^2} \int_{\mathbbm C}\d^2\lambda \,
\chi(\lambda,p)\,\e^{-(p-q)|\lambda|^2/2}\,\e^{\alpha\lambda^*-\alpha^*\lambda}\notag\\[1ex]
&=
\frac{1}{\pi^2} \int_{\mathbbm C}\d^2\lambda \,
\e^{-(p-q)|\lambda|^2/2}\,\e^{\alpha\lambda^*-\alpha^*\lambda}\nonumber\\
&\hspace{2.5cm}\times \underbrace{
\int_{\mathbbm C}\d^2\beta \, W(\beta,p)\,\e^{\lambda^*\beta-\lambda\beta^*}
}_{\displaystyle \chi(\lambda,p)}\notag\\[1ex]
&=\frac{2}{\pi(p-q)} \int_{\mathbbm C}\d^2\beta \, W(\beta,p)\,
\exp\left[-\frac{2|\alpha-\beta|^2}{(p-q)}\right]\,.
\end{align}
Equation~(\ref{W:q:p}) allows to pass from one ordering to the other. Note that as $q$ decreases the peaks in the quasi-probability distribution become broader.
\par
We have seen that there is a strict relation between the characteristic functions and the quasi-probability distributions. However, $W(\alpha,p)$ can be obtained directly from the density operator $\hrho$, without passing through the characteristic function formalism, namely \cite{cahill}:
\begin{equation}
W(\alpha,p)=\mbox{Tr}[\hrho\, \hD(\alpha)\,\hT(p)\,\hD^{\dag}(\alpha)], \quad (p<1)
\end{equation}
where:
\begin{equation}
\hT(p) = \frac{2}{\pi(1-p)}\, :\! \exp\left(-\frac{2}{1-p}\, \ha^{\dag}\ha \right)\!\!: \,,
\end{equation}
therefore we have:
\begin{equation}\label{W:p:def:2}
W(\alpha,p) = \frac{2}{\pi(1-p)}\sum_{n=0}^{\infty} \left(-\frac{1+p}{1-p}\right)^n
\bra{n} \hD^{\dag}(\alpha)\,\hrho\,\hD(\alpha)\ket{n}.
\end{equation}
It is worth noting that $\bra{n} \hD^{\dag}(\alpha)\,\hrho\,\hD(\alpha)\ket{n}$ is the photon number distribution of the state $\hrho$ after a displacement of amount $-\alpha$, and it can be also retrieved experimentally, thus allowing to reconstruct the $p$-ordered Wigner function.

\subsection{Glauber-Sudarshan $P$-representation} The so-called $P$-representation allows to write a density operator as:
\begin{equation}\label{P:rep}
\hrho  = \int_{\mathbbm C}\d^2\alpha\, P(\alpha)\, \ket{\alpha}\bra{\alpha},
\end{equation}
where $\ket{\alpha}$ are coherent states and $P(\alpha)$ is called $P$-function. If we put $p=1$ into Eq.~(\ref{W:p:def}) we obtain:
\begin{equation}
W(\alpha,1) = P(\alpha).
\end{equation}
\par
To calculate $P(\alpha)$ given $\hrho$, we can start from the identity:
\begin{equation}
\langle - \beta | \hrho | \beta \rangle = \int_{\mathbbm C}\d^2\alpha\, P(\alpha)\,
e^{-|\alpha|^2 - |\beta|^2 + \alpha^* \beta - \alpha \beta^*},
\end{equation}
and, defining the Fourier transform as:
\begin{equation}
g(\beta) = \int_{\mathbbm C}\d^2\alpha\, f(\alpha)\, e^{ \alpha^* \beta - \alpha \beta^*}
\end{equation}
by applying the inverse Fourier transform, we get the useful formula:
\begin{equation}
P(\alpha) = \frac{e^{|\alpha|^2}}{\pi^2} \int_{\mathbbm C}\d^2\beta\, \langle - \beta | \hrho | \beta \rangle\,
e^{|\beta|^2 + \alpha \beta^* - \alpha^* \beta}.
\end{equation}
By expanding $\hrho$ on the Fock basis,  $\hrho = \sum_{nm} \varrho_{nm}| n\rangle \langle m |$, we obtain the formal expression of the $P$-function \cite{agarwal}:
\begin{equation}\label{P:Dd:exp}
P(\alpha) = \frac{e^{|\alpha|^2}}{\pi^2} \sum_{nm} \varrho_{nm}\, \frac{(-1)^{n+m}}{\sqrt{n!\, m!}} \,
\frac{\partial^{n+m}}{\partial \alpha^n\, \partial (\alpha^*)^m} \, \delta^{(2)}(\alpha),
\end{equation}
that clearly shows its singular nature in general.

\subsection{The Wigner function} The Wigner function  $W(\alpha)$, introduced by E.~P.~Wigner, is given by Eq.~(\ref{W:p:def}) with $p=0$, namely, $W(\alpha,0) = W(\alpha)$. Furthermore, we can write:
\begin{equation}
W(\alpha)=\frac{2}{\pi}\, \mbox{Tr}[\hrho\, \hD(\alpha)\,\hat{\Pi}(p)\,\hD^{\dag}(\alpha)],
\end{equation}
where we introduced the parity operator $\hat{\Pi} = (-1)^{\ha^{\dag}\ha}$. Note that
$\hD(\alpha)\,\hat{\Pi}(p)\,\hD^{\dag}(\alpha) = \hD(2\alpha)\,\hat{\Pi}(p) =
\hat{\Pi}(p)\,\hD^{\dag}(2\alpha)$

\subsection{The Husimi or $Q$-function} The $Q$-function follows from Eq.~(\ref{W:p:def:2}) when $p=-1$:
\begin{equation}
Q(\alpha) = W(\alpha,0) = \frac{1}{\pi} \bra{\alpha} \hrho \ket{\alpha}. 
\end{equation}
Note that in this case we really have a probability distribution, since $Q(\alpha)\ge 0$ and normalized.

\begin{figure}[t]
\begin{center}
    \includegraphics[width=0.55\columnwidth]{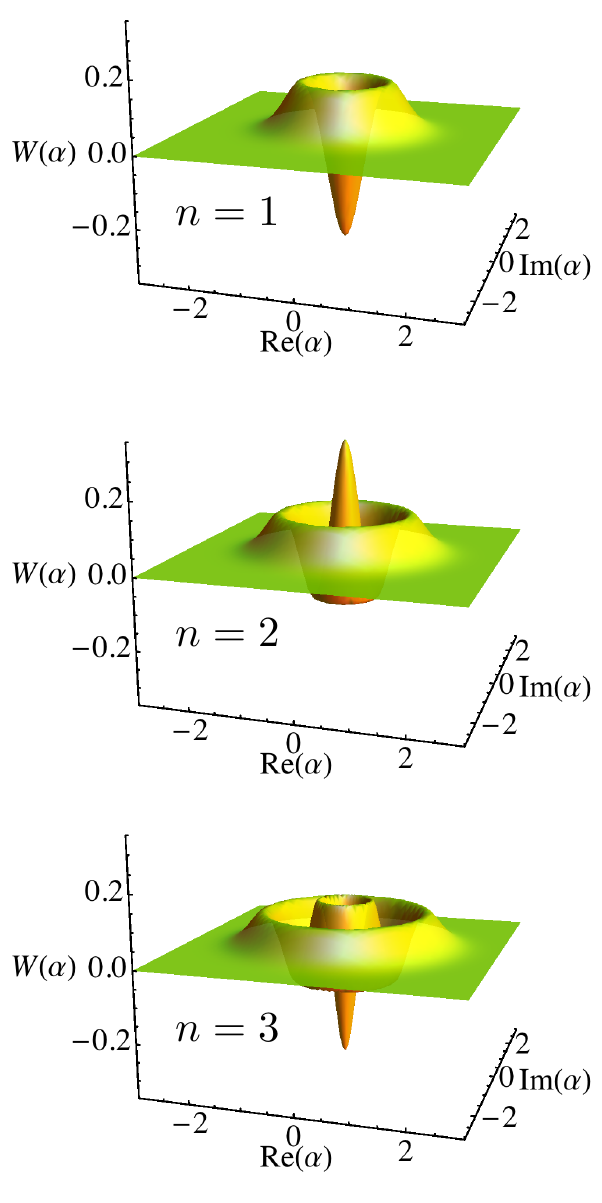}
    \vspace{-0.3cm}
\caption{\label{f:Wfock} Wigner function of the Fock state $\ket{n}$ for three values of $n$. Note that Wigner functions may have negative values.}
\end{center}
\end{figure}

\begin{figure}[t]
\begin{center}
    \includegraphics[width=0.55\columnwidth]{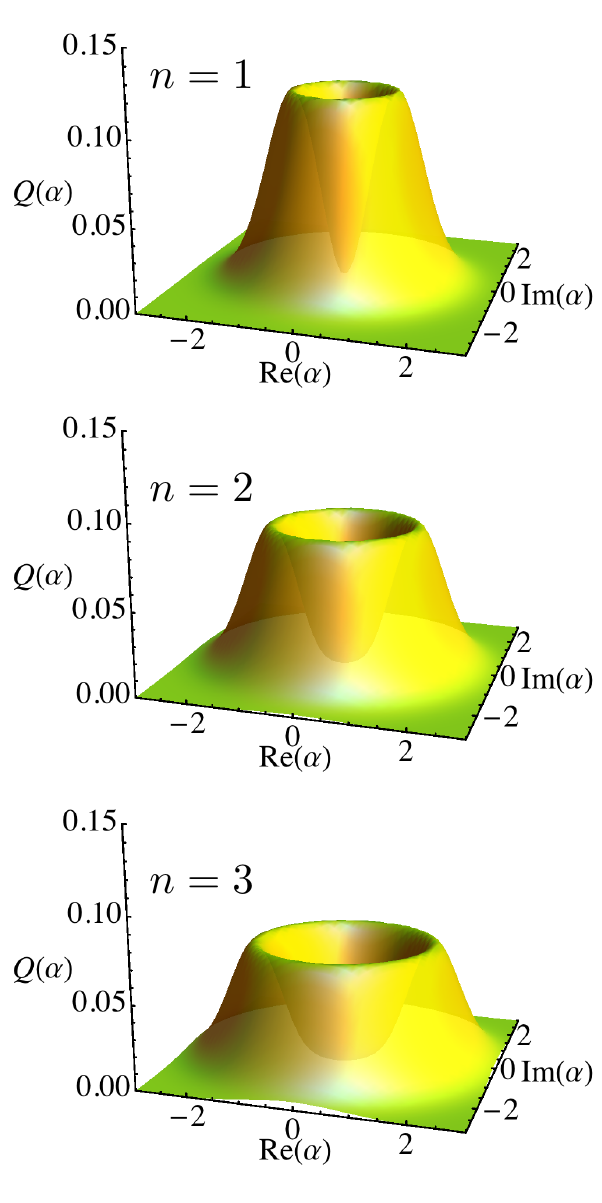}
    \vspace{-0.3cm}
\caption{\label{f:Qfock} $Q$-function of the Fock state $\ket{n}$ for three values of $n$. Note that, with respect to the Wigner functions in Fig.~\ref{f:Wfock}, the corresponding $Q$-functions are always positive.}
\end{center}
\end{figure}

\begin{example}\label{fock:ex}
If we consider a Fock state $\ket{n}$ we have:
\begin{align*}
&W(\alpha,1) = P(\alpha) = \frac{e^{|\alpha|^2}}{n!}
\frac{\partial^{2n}}{\partial\alpha^n\, \partial (\alpha^*)^n} \, \delta^{(2)}(\alpha),
\\[1ex]
&W(\alpha,0) = W(\alpha) = \frac{1}{\pi}\, (-1)^{n}\, \e^{-2|\alpha|^2}\, L_n(4|\alpha|^2),
\\[1ex]
&W(\alpha,-1) = Q(\alpha) = \frac{1}{\pi}\, \e^{-|\alpha|^2}\, \frac{|\alpha|^{2n}}{n!}.
\end{align*}
It is clear that decreasing the value of $p$ from $1$ to $-1$ the corresponding $W(\alpha,p)$ becomes more and more regular. In Figs.~\ref{f:Wfock} and \ref{f:Qfock} we plot the Wigner function and the $Q$-function, respectively, of a Fock state $\ket{n}$ for different values of $n$. $ $ \hfill $\blacksquare$
\end{example}


\begin{figure}[t]
\begin{center}
    \includegraphics[width=0.55\columnwidth]{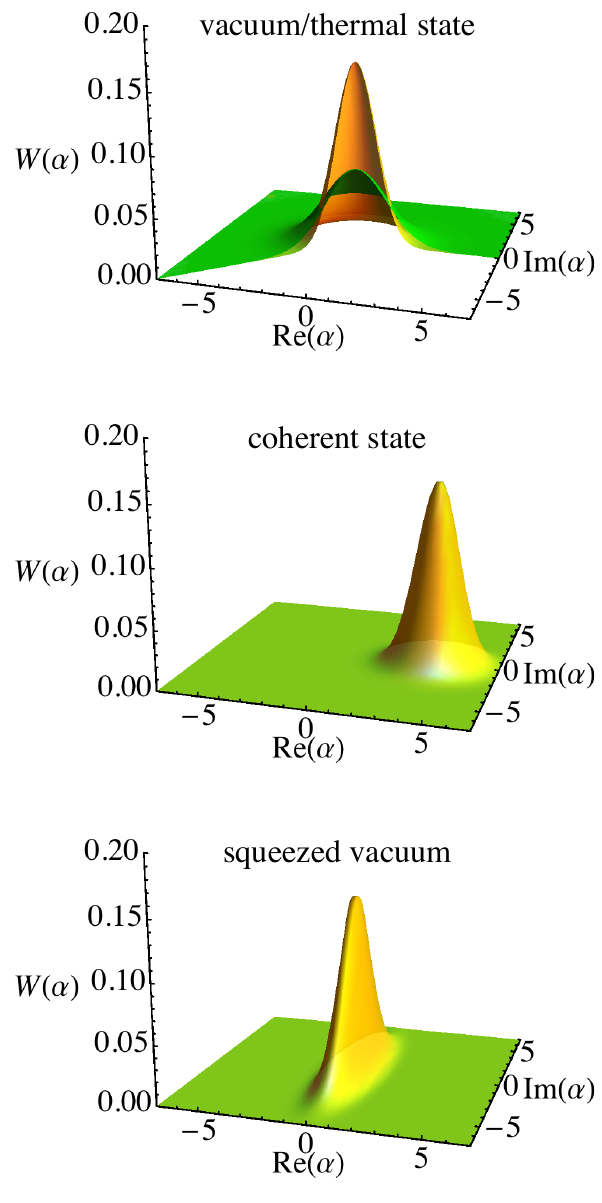}
    \vspace{-0.3cm}
\caption{\label{f:Wgs} Wigner functions of the vacuum state (left plot, yellow) and of a thermal state (left plot, green), of a coherent state (center) and of a squeezed vacuum state (right). Note that, though the squeezed vacuum state is nonclassical, its Wigner function if positive definite.}
\end{center}
\end{figure}

As in the case of the characteristic function, given the Wigner function $W(\alpha)$ of the state $\hrho$ we have:
\begin{equation}
\hD(\beta)\,\hrho\,\hD^{\dag}(\beta) \to W(\alpha-\beta),
\end{equation}
and
\begin{equation}
 \hS(r)\,\hrho\,\hS^{\dag}(r) \to W(\alpha\, \cosh r - \alpha^* \sinh r).
\end{equation}
In Fig.~\ref{f:Wgs} we plot the Wigner functions of the vacuum, the thermal, the coherent and of the squeezed vacuum states, respectively.

\subsection{Trace rule for the Wigner functions} Given two operators $\hA$ and $\hB$ and the corresponding Wigner functions $W_{\hA}(\alpha)$ and $W_{\hB}(\alpha)$, we have:
\begin{equation}
\mbox{Tr}[\hA\hB] = \pi \int_{\mathbbm C} \d^2\alpha\,
W_{\hA}(\alpha)\,W_{\hB}(\alpha).
\end{equation}

\section{Nonclassical states}\label{s:NCstates}
Albeit a thorough discussion about the nonclassicality is beyond the aim of these pages, in this section we draw some useful comment starting from the example~\ref{fock:ex}, where, in particular, we calculated the $P$-function and the Wigner function of the Fock states. In this context, we focus on a criterion for nonclassicality stemming from physical constraints and leave to the interested reader the criteria funded on information theoretic concepts (see, for instance, \cite{AP:ineq}).

We have seen in sect.~\ref{s:fano}, how the Fock states are sub-Poissonian (reaching $\cF=0$, the minimum value of the Fano factor) and, therefore, cannot be described classically: they are nonclassical. As a consequence, the results of the example~\ref{fock:ex} show that Fock states have a highly singular $P$-function (a sum of derivatives of Dirac's delta functions): only {\it classical} states can be represented as a {\it mixture} of coherent states, namely, the function $P(\alpha)$ in Eq.~(\ref{P:rep}) is a probability distribution (positive semidefinite and normalized). Note that, for a coherent state $\hrho  = \ket{\beta}\bra{\beta}$, we find $P(\alpha) = \delta^{(2)}(\alpha - \beta)$, that can be seen as the ``boundary'' between classical and nonclassical states. Therefore, in the presence of nonclassical states, such as the Fock states or the squeezed states, the corresponding $P$-function is more singular than a Dirac's delta: this can be easily seen from the general expansion Eq.~(\ref{P:Dd:exp}).

Let us now turn the attention on the Wigner functions. Still considering the example~\ref{fock:ex} and Fig.~\ref{f:Wfock}, we find that the Wigner functions of the Fock states have negative values. Also the negativity of the Wigner function, like the sub-Poissonian features of a state, is a {\it sufficient} signature of its nonclassicality: if  its $P$-function is a proper probability distribution, the Wigner function is positive definite, being linked through Eq.~(\ref{W:q:p}) with $q=0$ (Wigner function) and $p=1$ ($P$-function). Nevertheless, if the Wigner function is positive definite, the corresponding $P$-representation could be also highly singular and, thus, the state is nonclassical. This is the case of the squeezed states. 

For the sake of simplicity here we consider the squeezed vacuum, whose $p$-ordered characteristic function reads (without lack of generality we assume $r \in {\mathbb R}$, $r > 0$):
\begin{equation}\label{p:ord:sq}
\chi_{\rm sq}(\lambda,p) = \exp\left\{
-  \frac{e^{2r} - p}{2}\, \lambda_{\Re}^2 -  \frac{e^{-2r} - p}{2}\, \lambda_{\Im}^2
\right\},
\end{equation}
where $\lambda = \lambda_{\Re} + i \lambda_{\Im}$. Therefore, $\chi_{\rm sq}(\lambda,p)$ is unbounded if $p > e^{-2r}$ and the $P$-function, that we obtain by inserting $\chi_{\rm sq}(\lambda,p)$ into Eq.~(\ref{W:p:def}) with $p=1$, can be expressed only in terms of the derivatives of the Dirac's delta exploiting, for example, Eq.~(\ref{P:Dd:exp}). If $p < e^{-2r}$, Eq.~(\ref{p:ord:sq}) represents a two-dimensional Gaussian function and, thus, also the Wigner function $W(\alpha, 0)$ is a well-defined Gaussian (see Fig.~\ref{f:Wgs}, bottom panel) \cite{olivares}.
\newline\indent
Albeit on this section we considered only the nonclassical aspects, we stress that states endowed with negative Wigner functions or, more in general, with non-Gaussian features, are also relevant resources for the development of continuous-variable quantum information technologies \cite{albarelli,grosshans}.
\subsection{On the generation of nonclassical states}
\begin{figure}[t]
\begin{center}
    \includegraphics[width=0.75\columnwidth]{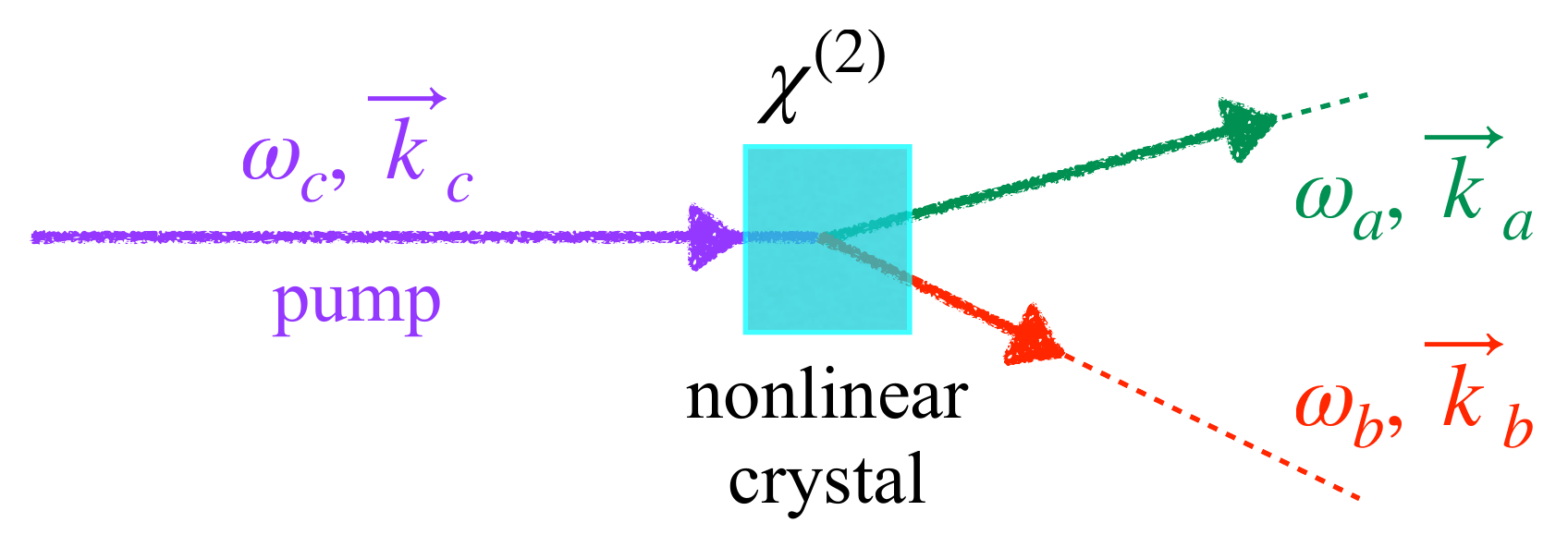}
    \vspace{-0.3cm}
\caption{\label{f:PDC} Scheme of the down-conversion process leading to single- and two-mode squeezing.}
\end{center}
\end{figure}

In sections \ref{s:smSq} and \ref{s:tmSq} we have introduced the Hamiltonians leading to single- and two-mode squeezing, respectively. These Hamiltonians are implemented through a three-wave interaction on non-linear crystal characterized by a second order nonlinear susceptibility $\chi^{(2)}$: thanks to the nonlinearity, one photon of the input mode $\hc$, the ``pump'', is converted into two photons of the  modes, $\ha$ and $\hb$, respectively (see Fig.~\ref{f:PDC}). The frequencies and the directions of the output modes should fulfil the {\it phase-matching conditions}, namely,  $\omega_c = \omega_a + \omega_b$ (frequency conservation) and $\vec{k}_c = \vec{k}_a + \vec{k}_b$ (momentum conservation). Being $\omega_c > \omega_a, \omega_b$, the process is called ``down-conversion''. Usually, the pump is a high intensity laser beam and it is considered as a classical field whose complex amplitude remains constant (parametric process), leading to the squeezing operators described in sections \ref{s:smSq}, for $\omega_a = \omega_b$ and $\vec{k}_a = \vec{k}_b$ and \ref{s:tmSq}, otherwise. In addition, the process of the parametric down-conversion is at the basis of the techniques to generate entangled photons \cite{kwiat} that can be exploited for advanced applications.

Nonclassical fields can be also generated in the presence of third-order order nonlinearities (Kerr effect): now, {\it two} photons of the pump are converted into the two new photons (four-wave mixing). This process is of particular interest for the generation of nonclassical fields at telecomm wavelength and has been studied in optical fibres to obtain squeezing, too \cite{bachor}.

Another interesting and effective way to create and manipulate quantum states exploits the linear optical elements and the photodetection we introduced in the previous sections. In this scenario, one can add or subtract single photons to an input state obtaining a new one with enhanced nonclassical properties that can be used for fundamental tests of quantum mechanics or for applications in quantum technologies \cite{zavatta,allevi,olivares05}.

We finally recall that nonclassical optical states, such as the Schr\"odinger's cat states, may be obtained in the framework of cavity quantum electrodynamics, through the interaction between atoms and electromagnetic fields inside microwave cavities \cite{haroche}: in this case the optical states are ``trapped'' inside the finite volume of the cavity and they are not ``travelling'' as in the case discussed above.

\section{Wigner function and quadrature expectation values}\label{s:Wexp}
If we introduce the quadrature $\hx = \ha+\ha^{\dag}$ with eigenvectors $\ket{x}$, namely, $\hx\ket{x} = x \ket{x}$, the generic quadrature $\hx_\theta$  can be written as:
\begin{align}
\hx_{\theta} = \ha\,\e^{-i\theta}+\ha^{\dag}\,\e^{i\theta} \equiv \hU_{\theta}^{\dag}\, \hx\, \hU_{\theta},
\end{align}
where $\hU_{\theta} = \exp(-i\theta \, \ha^{\dag}\ha)$ is a phase-shift operator. If $\hx_\theta \ket{x}_\theta = x_\theta \ket{x}_\theta$, then we have:
\begin{equation}
\underbrace{\hU_{\theta}^{\dag}\, \hx\, \hU_{\theta}}_{\displaystyle \hx_\theta} \ket{x}_\theta =
x_\theta \ket{x}_\theta,
\end{equation}
that is:
\begin{equation}
\hx\, \underbrace{\hU_{\theta} \ket{x}_\theta}_{\displaystyle \ket{x_\theta}} =
x_\theta \underbrace{\hU_{\theta} \ket{x}_\theta}_{\displaystyle \ket{x_\theta}}, \quad
\Rightarrow \quad \hx \ket{x_\theta} = x_\theta \ket{x_\theta},
\end{equation}
that is the state $\ket{x_\theta} = \hU_\theta\ket{x}_\theta$ is eigenstate of $\hx$ with eigenvalue $x_\theta$. Therefore, the probability $p(x;\theta)$ to obtain an outcome $x$ measuring the quadrature $\hx_\theta$ given the state $\hrho$ can be written as:
\begin{equation}\label{p:x:theta}
p(x;\theta) = \bra{x}\hU_\theta\,\hrho\,\hU_\theta^{\dag} \ket{x},
\end{equation}
since $\hU_\theta^{\dag} \ket{x}$ is eigenstate of $\hx_\theta$ with eigenvalue $x$.
\par
In this section we will show how the Wigner function $W(\alpha)$ of the state $\hrho$ is related to the probability $p(x;\theta)$. First of all we should rewrite the characteristic function $\chi(\lambda)$ of $\hrho$ in cartesian notation, that is, we write $\lambda = u + i v$, where, $u,v \in {\mathbbm R}$. We have:
\begin{equation}
\chi(\lambda) = \mbox{Tr}[\hrho\, \e^{\lambda\ha^\dag-\lambda^*\ha}] =
\mbox{Tr}[\hrho\, \e^{i(v\, \hx-u\, \hy)}],
\end{equation}
where $\hy = i(\ha^\dag - \ha)$. Now, using Theorem~\ref{thm:exp:sum} and evaluating the trace on the basis $\{\ket{x}\}$ of the eigenstates of $\hx$ we have:
\begin{equation}
\mbox{Tr}[\hrho\, \e^{iv\, \hx}\,\e^{-iu\, \hy}]\, \e^{-iuv} =
\int_{\mathbbm R}\d x\, \bra{x}  \hrho\, \e^{iv\, \hx}\,\e^{-iu\, \hy} \ket{x}\, \e^{-iuv}.
\end{equation}
Since $[\hx,\hy]=2i$, we have $\e^{-iu\, \hy} \ket{x} = \ket{x+2u}$, thus we obtain:
\begin{equation}
\int_{\mathbbm R} \d x\, \bra{x}  \hrho\, \e^{iv\, \hx} \ket{x + 2u}\, \e^{-iuv}  =
\int_{\mathbbm R} \d q\, \bra{q-u}  \hrho \ket{q + u}\, \e^{ivq} ,
\end{equation}
where we used the change of variable $x = q-u$. Summarizing, we can write the characteristic function as:
\begin{equation}
\chi(u,v) = \int_{\mathbbm R} \d q\, \bra{q-u}  \hrho \ket{q + u}\, \e^{ivq}.
\end{equation}
The corresponding Wigner function is given by Eq.~(\ref{W:p:def}). In order to write also $W(\alpha)$ in cartesian notation we should write $\alpha = (x+iy)/2$, where the factor $1/2$ is due to the definition of the quadrature operator. Since $\alpha\lambda^* - \alpha^*\lambda = i(uy-vx)$ we have:
\begin{align}
W(x,y) &= \frac{1}{(2\pi)^2} \int_{\mathbbm R} \d u \int_{\mathbbm R} \d v \,
\chi(u,v)\, \e^{i(uy-vx)} \\[1ex]
&= \frac{1}{2\pi} \int_{\mathbbm R} \d u 
\int_{\mathbbm R} \d q\,
\bra{q-u}  \hrho \ket{q + u}\, \e^{iuy}\, \notag\\[1ex]
&\hspace{2.5cm}\times\underbrace{
\frac{1}{2\pi} \int_{\mathbbm R} \d v \,\e^{iv(q-x)}
}_{\displaystyle \delta^{(2)}(q-x)}.
\end{align}
Therefore, after the integration over $q$, we obtain the original definition of the Wigner function:
\begin{equation}\label{Wigner:original}
W(x,y)= \frac{1}{2\pi} \int_{\mathbbm R} \d u \,
\bra{x-u}  \hrho \ket{x + u}\, \e^{iuy}.
\end{equation}
It is now straightforward to see that:
\begin{equation}
p(x) = \int_{\mathbbm R} \d y\, W(x,y) \equiv \bra{x}\hrho\ket{x},
\end{equation}
or, more in general, since the Wigner associated with the state $\hU_\theta\,\hrho\,\hU_\theta^{\dag}$ is:
\begin{equation}
\hU_\theta\,\hrho\,\hU_\theta^{\dag} \to W(x\cos\theta-y\sin\theta,y\cos\theta+x\sin\theta),
\end{equation}
we obtain that the probability distribution of the outcomes of the generic quadrature $\hx_\theta$ is given by the following marginal of the suitably transformed Wigner function \cite{leonhardt}:
\begin{align}
p(x;\theta) &= \int_{\mathbbm R} \d y\, W(x\cos\theta-y\sin\theta,y\cos\theta+x\sin\theta)
\notag\\[1ex]
&\equiv \bra{x}\hU_\theta\,\hrho\,\hU_\theta^{\dag} \ket{x}.
\end{align}
\par
\begin{figure}[t]
\begin{center}
    \includegraphics[width=0.7\columnwidth]{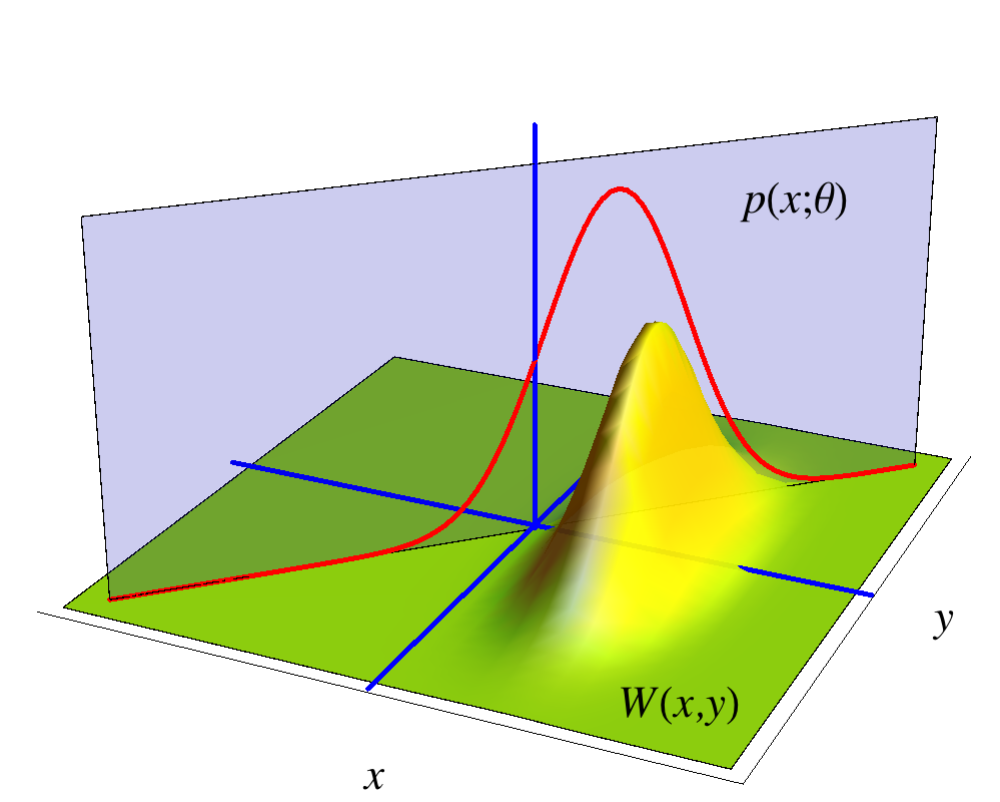}
\caption{\label{f:Wp} Winger function of a displaced squeezed vacuum state (yellow surface) and its marginal $p(x;\theta)$ (red line).}
\end{center}
\end{figure}
In Fig.~\ref{f:Wp} we show the Wigner function $W(x,y)$ of a displaced squeezed vacuum state and its marginal $p(x;\theta)$.

\section{Measuring the expectations of the quadrature operator}\label{s:QOexp}
In this section we describe  the \emph{balanced homodyne detection}, a method to measure the quadrature operator \cite{FOP}.
We have seen that when two modes $\ha$ and $\hb$ interact through a balanced BS their evolution is given by Eqs.~(\ref{BS:heisenberg}) and we can write the outgoing modes $\hc$ and $\hd$ as:
\begin{equation}
\hc = \frac{\ha + \hb}{\sqrt{2}}, \quad \mbox{and} \quad
\hd = \frac{\hb - \ha}{\sqrt{2}}.
\end{equation}
If we define the sum and difference output photocurrent $\hI_{\pm} = \hc^\dag\hc \pm \hd^\dag\hd$ we have that $\hI_{+} = \ha^\dag\ha + \hb^\dag\hb$ (energy is conserved) and:
\begin{align}
\hI_{-} &= \hc^\dag\hc - \hd^\dag\hd, \\[1ex]
&= \hb^\dag\ha + \hb \ha^\dag\,.
\end{align}
If we now assume that the input state is the factorized state $\hrho\otimes \ket{z\,\e^{i\theta}}\bra{z\,\e^{i\theta}}$, where $\ket{z\,\e^{i\theta}}$ is a coherent state, $z\in {\mathbbm R}$, we obtain the following expectation:
\begin{equation}
\langle \hI_{-} \rangle = z\, \langle \hx_\theta \rangle
\quad \Rightarrow \quad
\langle \hx_\theta \rangle = \frac{\langle \hI_{-} \rangle}{z} =
\langle \hat{\cI} \rangle,
\end{equation}
where $\hx_\theta = \ha\, \e^{-i\theta} + \ha^\dag\,\e^{i\theta}$,
$\langle \hx_\theta \rangle = \mbox{Tr}[\hrho\,\hx_\theta]$ and $\hat{\cI} = (\hb^\dag\ha + \hb \ha^\dag)/z$.
Therefore, the mean value of the quadrature $\hx_\theta$ given the state $\hrho$ can be measured mixing the state at a balanced BS with a \emph{local oscillator} $\ket{z\,\e^{i\theta}}$, measuring the different photocurrent and normalizing the outcome with respect to $z$. On the other hand, if we evaluate the second moment $\langle \cI^2\rangle$, upon the substitution $\hb \to z\,\e^{i\theta}$, we find:
\begin{equation}
\hat{\cI}^2 = \hx_\theta^{2} + \frac{\ha^{\dag}\ha}{z^2},
\end{equation}
thus one should have $\langle \ha^{\dag}	\ha \rangle = \mbox{Tr}[\hrho\,\ha^{\dag}\ha]\ll z^2$ in order to obtain the actual quadrature moment, namely $\langle \hat{\cI}^2 \rangle \approx \langle \hx_\theta^{2} \rangle$.

\begin{figure}[t]
\begin{center}
    \includegraphics[width=0.5\columnwidth]{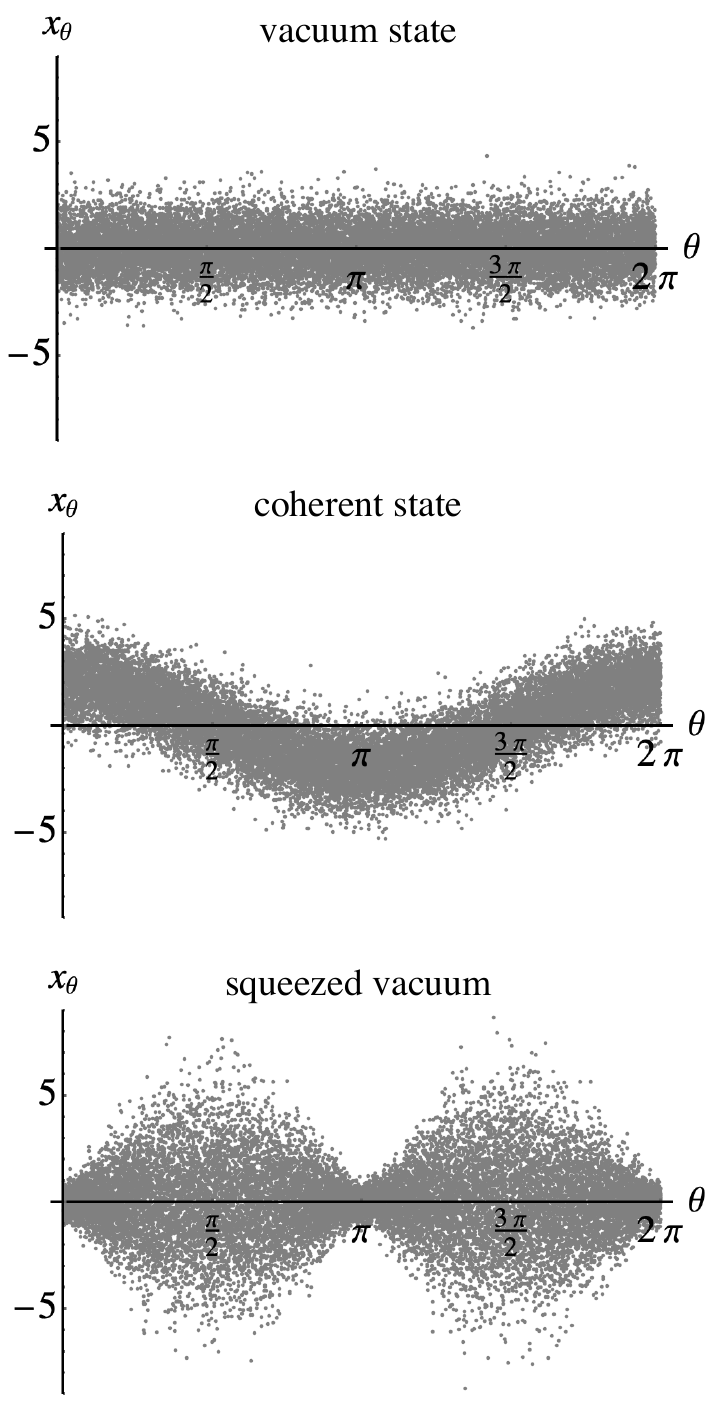}
    \vspace{-0.3cm}
\caption{\label{f:hd:trace} Monte Carlo simulations of homodyne traces for the vacuum state, a coherent state $\ket{\alpha}$, $\alpha=1$, and a squeezed vacuum $\hS(r)\ket{0}$ with $\langle \hn \rangle =1$ and $r<0$, corresponding to $7.7$~dB of squeezing.}
\end{center}
\end{figure}
In Fig.~\ref{f:hd:trace} we show the Monte Carlo simulations of homodyne traces in the case of the vacuum state, a coherent state and a squeezed vacuum state.

\section{Homodyne tomography}\label{s:tomo}
In this section we show how we can obtain the expectation of an observable $\hA$ given a state $\hrho$ and its homodyne data sample $\{ (\theta_k, x_k) \}$, $k=1,\ldots , M$, $x_k$ being the outcome from the observation of the quadrature $\hx_{\theta_k}$ \cite{tomography}. This is crucial to achieve the experimental reconstruction of optical states and their advanced analysis \cite{dauria, olivares:g2}. We remark that other approaches, such as the ones based on minimax estimation, may be also useful to retrieve the Wigner function of the addressed states in the presence of noisy data or low quantum detection efficiency \cite{butuchea, esposito}.
\par
Exploiting the Glauber formula, we can write the operator $\hA$ as:
\begin{equation}
\hA = \frac{1}{\pi} \int_{\mathbbm C} \d^2 \alpha\,
\mbox{Tr}\left[ \hA\, \hD(\alpha) \right] \hD^{\dag}(\alpha)
\end{equation}
or, writing $\alpha = i y \, \e^{-i\theta}$, as:
\begin{equation}
\hA = \frac{1}{\pi} \int_{0}^{\pi}\d\theta \int_{\mathbbm R} \d y\,|y|\,
\mbox{Tr}\left[ \hA\, \e^{i y \hx_\theta} \right] \e^{-i y \hx_\theta}.
\end{equation}
The expectation $\langle \hA \rangle = \mbox{Tr}[\hrho\, \hA]$ is thus given by:
\begin{equation}
\langle \hA \rangle = 
\frac{1}{\pi} \int_{0}^{\pi}\d\theta \int_{\mathbbm R} \d y\,|y|\,
\mbox{Tr}\left[ \hA\, \e^{i y \hx_\theta} \right]
\mbox{Tr}\left[ \hrho\, \e^{-i y \hx_\theta} \right],
\end{equation}
and, evaluating the last trace using the basis $\{\ket{x}_\theta\}$ of the eigenvectors of the quadrature $\hx_\theta$, $\hx_\theta\ket{x}_\theta = x\ket{x}_\theta$, namely:
\begin{equation}
\mbox{Tr}\left[ \hrho\, \e^{-i y \hx_\theta} \right] =
\int_{\mathbbm R} \d x\,
\underbrace{
{}_{\theta} \bra{x} \hrho \ket{x}_{\theta}}_{\displaystyle p(x;\theta)}
\, \e^{-i y x},
\end{equation}
we have:
\begin{align}
\langle \hA \rangle &= 
\frac{1}{\pi} \int_{0}^{\pi}\d\theta \int_{\mathbbm R} \d x\,
p(x;\theta)\, {\mathcal R}[\hA](x,\theta)\\[1ex]
&\equiv \overline{{\mathcal R}[\hA](x,\theta)}\, ,
\label{tomography}
\end{align}
where we introduced the estimator of the operator ensemble average $\langle \hA \rangle$ given by:
\begin{equation}\label{HD:est}
{\mathcal R}[\hA](x,\theta) = \int_{\mathbbm R} \d y\,|y|\, 
\mbox{Tr}\left[ \hA\, \e^{i y (\hx_\theta - x)} \right].
\end{equation}
Equation (\ref{tomography}) is at the basis of quantum homodyne tomography.
It is worth noting that since the tomographic measurement is given in terms of the average $\overline{{\mathcal R}[\hA](x,\theta)}$ the precision of the measurement is given by the quantity (here we consider only the case of $\overline{{\mathcal R}[\hA](x,\theta)} \in {\mathbbm R}$):
\begin{equation}
\delta^2 = \overline{{\mathcal R}^2[\hA](x,\theta)} -
\left\{ \overline{{\mathcal R}[\hA](x,\theta)} \right\}^2,
\end{equation}
where:
\begin{equation}
\overline{{\mathcal R}^2[\hA](x,\theta)} = 
\frac{1}{\pi} \int_{0}^{\pi}\d\theta \int_{\mathbbm R} \d x\,
p(x;\theta)\, {\mathcal R}^2[\hA](x,\theta) .
\end{equation}
\par
In the case of the homodyne data sample $\{ (\theta_k, x_k) \}$, $k=1,\ldots , M$, where $\theta_k$ uniformly spans the interval $[0,\pi]$ and $M \gg 1$, Eq.~(\ref{tomography}) can be written as:
\begin{equation}
\langle \hA \rangle = \lim_{M \to \infty}
\frac{1}{M }\sum_{k=1}^{M} {\mathcal R}[\hA](x_k,\theta_k).
\end{equation}
Indeed, the finite sum:
\begin{equation}\label{tomography:sample}
\langle \hA \rangle =
\frac{1}{M }\sum_{k=1}^{M} {\mathcal R}[\hA](x_k,\theta_k)
\end{equation}
gives an approximation of the actual value of $\langle \hA \rangle$. Furthermore, using the central limit theorem, one finds that the tomographic estimation converges with a statistical error that decreases as $1/\sqrt{M}$.
\par
In the case of optical systems, one can reduce the estimation of the expectation of any operator to the estimation of normally ordered products of annihilation and creation operators $(\ha^{\dag})^n \ha^m$, whose estimator is:
\begin{equation}\label{R:tomo}
{\mathcal R}[(\ha^{\dag})^n \ha^m](x,\theta) = 
\e^{i (n-m) \theta}\, \frac{H_{n+m}(x/\sqrt{2})}{\sqrt{2^{n+m}}\, {n+m \choose n}},
\end{equation}
where $H_n(x)$ are Hermite polynomials.
\par
\begin{table}
\caption{\label{t:estimator} Estimator ${\mathcal R}_\eta[\hA](x,\theta)$ for some operators $\hA$.}
\renewcommand{\arraystretch}{2}
\begin{tabular}{ l l }
  \hline                       
$\hA$ \hspace{1.5cm} & ${\mathcal R}_\eta[\hA](x,\theta)$ \vspace{0.1cm}\\
  \hline  
$\ha$ & $\e^{ i \theta} x$ 	\\
$\ha^2$ & $\displaystyle \e^{ 2 i \theta} \left( x^2 -\frac{1}{\eta} \right)$ \\
$\hx_{\phi}$ & $2 x \cos(\theta - \phi)$ 	\\
	$\hx_{\phi}^2$ & $\displaystyle \left(x^2 -\frac{1}{\eta}\right)
	\{1 + 2 \cos[2(\theta-\phi)]\} + 1 $ 	\\
$\ha^\dag \ha$ & $\displaystyle \frac12 \left( x^2 -\frac{1}{\eta}\right)$ \\ 
$(\ha^\dag \ha)^2$ & $\displaystyle \frac{x^4}{6} - \left(\frac{2 - \eta}{2 \eta}\right) x^2 + \frac{1-\eta}{2 \eta^2}$
 \vspace{0.2cm}\\
 \hline  \\
\end{tabular}
\end{table}
Since quantum tomography has a practical utility, we note that in the presence of homodyne detection with non-unit quantum efficiency $\eta$, the homodyne probability is given by the convolution:
\begin{equation}
p_\eta(x;\theta) = \frac{1}{\sqrt{2 \pi \delta_\eta^2}}
\int_{\mathbbm R} \d x' \, p(x';\theta)\, 
\exp\left[- \frac{(x' - x )^2}{2 \delta_\eta^2}\right]
\end{equation}
with $\delta_\eta^2 = (1-\eta)/\eta$. In turn, the function in Eq.~(\ref{R:tomo}) should be replaced by:
\begin{equation}
{\mathcal R}_\eta[(\ha^{\dag})^n \ha^m](x,\theta) = 
\e^{i (n-m) \theta}\, \frac{H_{n+m}(\sqrt{\eta}x/\sqrt{2})}{\sqrt{(2\eta)^{n+m}}\, {n+m \choose n}}\, .
\end{equation}
In Table~\ref{t:estimator} we report the analytical expression of the estimator ${\mathcal R}_\eta[\hA](x,\theta)$ for some relevant operators $\hA$.

\section{Conclusions}\label{s:concl}
In these pages I have summarized the basic theoretical tools to deal with generation, manipulation and characterization of optical quantum states. Indeed, they didn't cover all the aspects of this interesting topics: this is far beyond the scope of this work (the interested reader can found further information in the essential list of references). Nevertheless, I hope that this introduction has been a useful tool both for the student that was looking for an ``advanced summary'' of quantum optics as well as for the researcher that applies it to describe and investigate our world.

\section*{Acknowledgments}
I would like to acknowledge A.~Allevi, M.~Bondani, F.~Castelli, S.~Cialdi, M.~Frigerio and M.~G.~A.~Paris for useful discussions and suggestions. This work has been partially supported by MAECI, Project No.~PGR06314 ``ENYGMA'' and by University of Milan, Project No.~\mbox{RV-PSR-SOE-2020-SOLIV} ``\mbox{S-O}~PhoQuLis''.

\end{document}